\documentclass[12pt,preprint]{aastex}

\begin{document}
\bibliographystyle{apj}

%  LaTeX definitions 
%
\def\slantfrac#1#2{\hbox{$\,^{#1}\!/_{#2}$}}
\def\figsize{\epsfxsize=1.0\columnwidth}
\def\ltsim{\raisebox{-.5ex}{$\;\stackrel{<}{\sim}\;$}}
\def\gtsim{\raisebox{-.5ex}{$\;\stackrel{>}{\sim}\;$}}
\def\hi{H\,{\sc i}}
\def\hii{H\,{\sc ii}}
\def\hei{He\,{\sc i}}
\def\heii{He\,{\sc ii}}
% end LaTeX definitions
%
\title{Low Temperature Opacities}

\author{Jason W. Ferguson\altaffilmark{1}, David R. Alexander\altaffilmark{1}, 
France Allard\altaffilmark{2}, 
Travis Barman\altaffilmark{1},
Julia G. Bodnarik\altaffilmark{1,3},
Peter H. Hauschildt\altaffilmark{4},
Amanda Heffner-Wong\altaffilmark{1},
Akemi Tamanai\altaffilmark{1,5} }

\altaffiltext{1}{Department of Physics, Wichita State University, Wichita, KS 67260-0032; 
jason.ferguson@wichita.edu; david.alexander@wichita.edu; travis.barman@wichita.edu}
\altaffiltext{2}{CRAL-ENS, 46 Allee d'Italie, Lyon, 69364 France, Cedex 07; fallard@ens-lyon.fr}
\altaffiltext{3}{Now at Gemini Observatory, 670 North A'ohoku Place, Hilo, HI 96720-2700; jbodnari@gemini.edu}
\altaffiltext{4}{Hamburger Sternwarte, Gojenbergsweg 112, 21029 Hamburg, Germany; yeti@hs.uni-hamburg.de}
\altaffiltext{5}{Now at Astrophysical Institute and University Observatory, 
Friedrich-Schiller-University Jena, Schillergaesschen 3, D-07745 Jena, Germany; akemi@astro.uni-jena.de }
 
\begin{abstract}

Previous computations of low temperature Rosseland and Planck mean opacities from 
\cite{af94a} are updated and expanded.  
The new computations include a more complete
equation of state with more grain species and updated optical constants.
Grains are now explicitly included in thermal equilibrium in the equation of state
calculation, which allows for a much wider range of grain compositions to be
accurately included than was previously the case. The inclusion of high temperature
condensates such as Al$_2$O$_3$ and CaTiO$_3$ significantly affects the total opacity over
a narrow range of temperatures before the appearance of the first silicate grains.

The new opacity tables
are tabulated for temperatures ranging from 30000~K to 500~K with gas densities from 10$^{-4}$~g~cm$^{-3}$ to 
10$^{-19}$~g~cm$^{-3}$.  Comparisons with previous Rosseland mean opacity calculations are discussed.  
At high temperatures, the agreement with OPAL and Opacity Project is quite good. Comparisons at
lower temperatures are more divergent as a result of differences in molecular and grain physics included in different 
calculations.  
The computation of Planck mean opacities performed with the opacity sampling method are shown
to require a very large number of opacity sampling wavelength points; previously published results obtained with 
fewer wavelength points are shown to be significantly in error.
Methods for requesting or obtaining the new tables are provided. 

\end{abstract}


% uncomment next line for preprint version
%\newpage

\section{Introduction}

%%%rewrote first sentence 
When modeling the transfer of radiation through optically thick material, it is often useful
to use pretabulated mean opacity tables, because performing computationally costly calculations 
of the frequency dependent opacity may not be feasible.  
There are many relevant applications in astrophysics for using mean opacities including 
the interiors of cool stars, giant planets, and disks of material forming stars and planets.

Recent opacity tables useful for modeling stellar interiors and envelopes 
come from two major sources, the OPAL opacities (\nocite{ir1991} \nocite{ir1993} \nocite{ir1996}
Iglesias \& Rogers (1991, 1993, 1996), \nocite{ri1992a} \nocite{ri1992b} 
Rogers \& Iglesias (1992a, 1992b) and \cite{rsi1996})and the Opacity Project
%%%removed parentheses 
\cite[OP hereafter]{OP1994}. The opacities from these two groups match very well at most temperatures, 
and their results have been 
%%% added cites for OP and OPAL resolutions
used to resolve several discrepancies between theoretical models and observations including 
evolutionary models of high-mass stars (\cite{SC1991}), intepretations of the HR diagram (\cite{chiosi1992}) and 
radiative acceleration in stellar envelopes (\cite{Gonzalez1995}).
Each of the above opacity databases are valid
for temperatures greater than 6000~K, but do not include the effects of 
molecules which become important at lower temperatures.

In cool gases molecules become dominating sources of opacity and must be included 
%%% removed important
in compilations of Rosseland opacity tables.  \cite{tsuji1966} first included the effects of 
molecular absorbers such as H$_2$O, CO, and OH.  The Wichita State University (WSU) low
temperature opacity group began publishing tables of opacities for temperatures
below 10000~K in 1975.  \cite{a75} computed opacities down 
to 700~K and included a crude estimate of the opacity due to dust grains.  
A better approximation to the dust opacity was included 
in \cite{ajr} and Alexander \& Ferguson (1994a, 1994b; AF94 hereafter)\nocite{af94a}\nocite{af94b}.  
\cite{sharp92} focused on molecular opacities and their application to accrection disks.
%%% added S03 to cite
Opacity tables for use in protoplanetary disk models were computed by \cite[S03 hereafter]{semenov2003} for
for gas and dust mixtures from 10~K to 10000~K.

%%% added ref. to phoenix... apparently forgotten in intitial writting!!
Table~1 illustrates the improvement in the WSU opacity calculations 
beginning in 1975 through the present.  Current opacities are computed with a modified version 
of the stellar atmosphere code {\sc Phoenix}.  Each of the changes indicated in the table will be discussed in 
one of the following sections starting with the equation of state followed by sources of opacity
data.  In Section~4 current opacities will be compared with AF94, 
%%% added S03 to list
OP, OPAL, and S03.  Section~5 focuses on the computation of Planck mean opacity values.
Methods of obtaining the new WSU opacity tables for the current set of compositions 
is described in Section~6 as well as how to request new or customized opacity tables.

\section{Equation of State}

As shown in Table~1 one of the most significant differences between the opacity tables 
of AF94 and current tables is within the equation of state (EOS hereafter).  
In AF94, only six grain species were included and their number densities were taken 
from pre-tabulated functions (precentage of grain material condensed) of temperature and 
pressure provided by Sharp (private communication).  Consequently, AF94 used two decoupled 
equations of state, one for gas with grains and one for pure gas.  While this procedure is 
reasonably accurate for compositions close to those provided to AF94 by Sharp (all of which have 
$X$ (hydrogen) and $Z$ (metal) abundances close to solar with differing C/O ratios), it becomes increasingly 
imprecise for compositions that deviate significantly from the solar mixture.  Tests show that for nearly
solar values of X the number abundances of grains in AF94 are consistent with the current work, however
%%% removed extreme X for hydrogen-poor
for a hydrogen poor gas differences do occur on the order of 20 times less grain material in AF94.  
The resulting mean opacity will subsequently be low by that factor, see below for a discussion of
the mean opacity calculation.

The current EOS contained in the {\sc phoenix} stellar atmosphere code most recently 
described in \cite{limitdust2001} contains 
several hundred molecules and dozens of solid and liquid species in chemical equilbrium.  
The new EOS represents a significant advance in the calculation of low temperature molecular 
and grain opacities when compared to AF94.  
The EOS of {\sc phoenix} is able to calculate the chemical equilibrium of 40 elements, 
including the ionization stages indicated in Table~2, along with the molecular 
and solid and liquid species for hundreds of species.  
%%% new sentence for ref
For the discussion in this paper we focus on molecular and solid phase species, since
liquids do not become abundant enough to affect the mean opacity.

%%% moved sentence to next paragraph
Figure~1 shows the abundances of condensates as a function of temperature for
a single gas pressure of 1~dyne~cm$^{-2}$ and a solar composition from \cite{gn93}.  
Several species have had their abundances summed for clarity.  
The silicate species
MgSiO$_3$, Mg$_2$SiO$_4$, and Fe$_2$SiO$_4$ are combined with the minor species
Na$_2$SiO$_3$ to make up the ``Silicates'' group.  The ``Fe'' 
group contains both Fe$-\alpha$ and Fe$-\gamma$ crystal structures.  The four crystal structures of 
Al$_2$O$_3$$-\alpha,\delta,\gamma,\kappa$ form the ``Al$_2$O$_3$'' group. 
The ``Titanates'' group contains CaTiO$_3$, MgTiO$_3$ and MgTi$_2$O$_5$.  
The ``Ca-Silicates'' contains CaMgSi$_2$O$_6$, Ca$_2$Al$_2$SiO$_7$ and 
Ca$_2$MgSi$_2$O$_7$. The ``Al-Silicates'' group contains Al$_6$Si$_2$O$_{13}$, Al$_2$SiO$_5$, 
KAlSi$_3$O$_8$ and NaAlSi$_3$O$_8$.  The choice of each of the groups is arbitrary and is
done only to make the figure more readable; such grouping is not done in the EOS or the 
opacity calculations of {\sc phoenix}.  

Figure~1 demonstrates the complicated nature of the equation of state for a cool gas.  
The first condensate to condense at the pressure shown in Fig.~1 is Al$_2$O$_3$ at about 1550~K
(although ZrO$_2$ condenses at a higher temperature its abundance is not large enough to be shown in 
the figure).  
Below 1400~K Al$_2$O$_3$ disappears in favor of the ``Ca-Silicates'' group
(which does contain a small amount of Al) and the ``Titanates''.
Just at and below 1200~K ``Silicates'' and the ``Fe'' condensates
form and dominate the abundance of grain material.  
%%% added "grain mixture"
The mean opacity of such a gas-grain mixture is discussed in Section~4.

\begin{figure}
\plotone{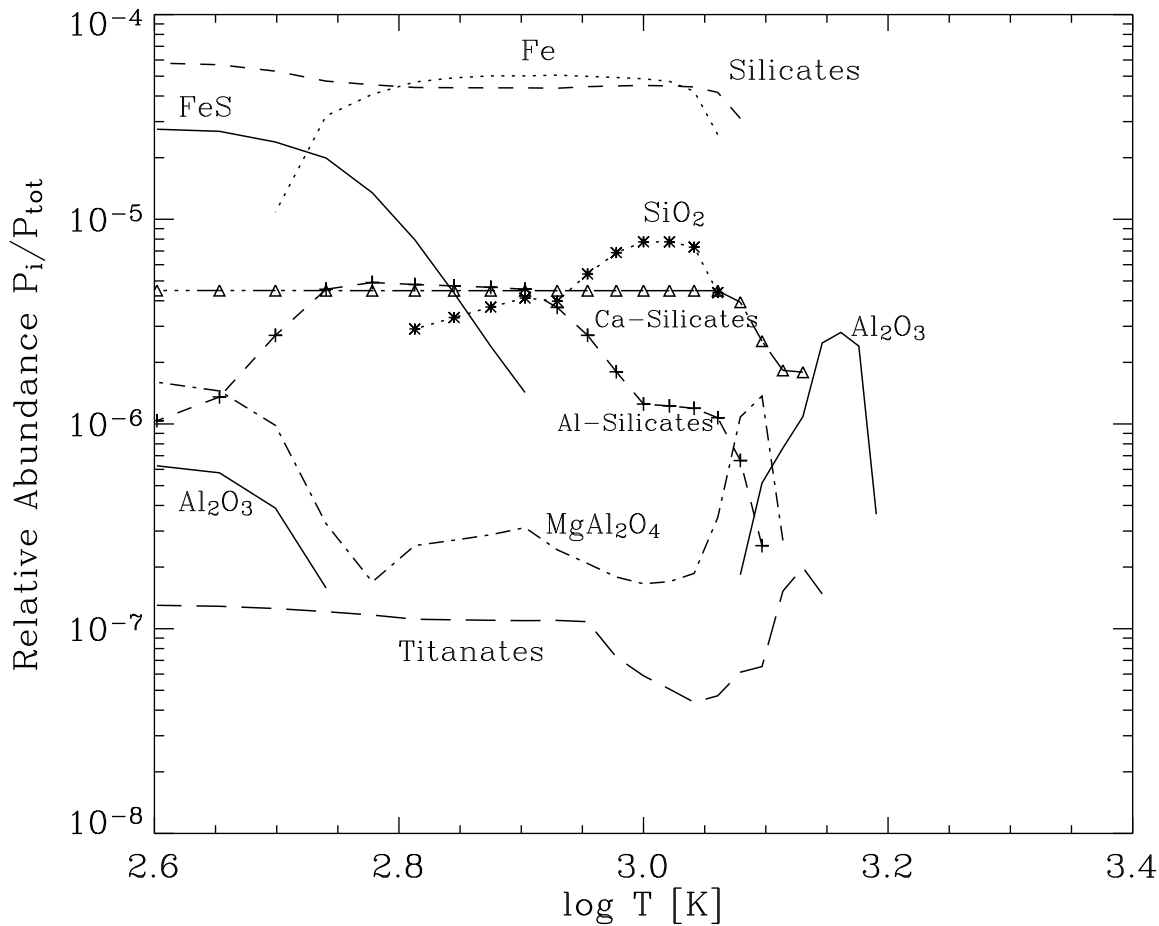}
\caption{The relative abundances of different groups of grain species for a single gas pressure
(1 dyne cm$^{-2}$) as a function of temperature.  
The choice of each group is arbitrary and is done only for clarity in this figure; 
the text describes the constituents of each group.}
\label{Figure 1}
\end{figure}

%%% cleaned up language
The interplay between solid species and gas-phase chemistry is illustrated by focusing 
on abundant titanium species in Figure~2.  
Two separate calculations of the EOS are shown in the figure, one with grains and one without (``nog'' in the figure).   
Both calculations assume solar abundances and a gas pressure of 1~dyne~cm$^{-2}$.  
When condensation is included in the 
EOS the abundances of the molecular species are greatly reduced in favor of solid species.  
For example, molecular TiO, which is an important absorber in cool stellar atmospheres disappears 
when grains are included in the EOS.  This removal will have a large impact on the opacity of the 
gas.  The effect of removing an important species such as titanium via (for example)
gravitational settling (see \cite{limitdust2001} and \cite{wh2004}) 
of solid particles can greatly affect the opacity.

\begin{figure}
\plotone{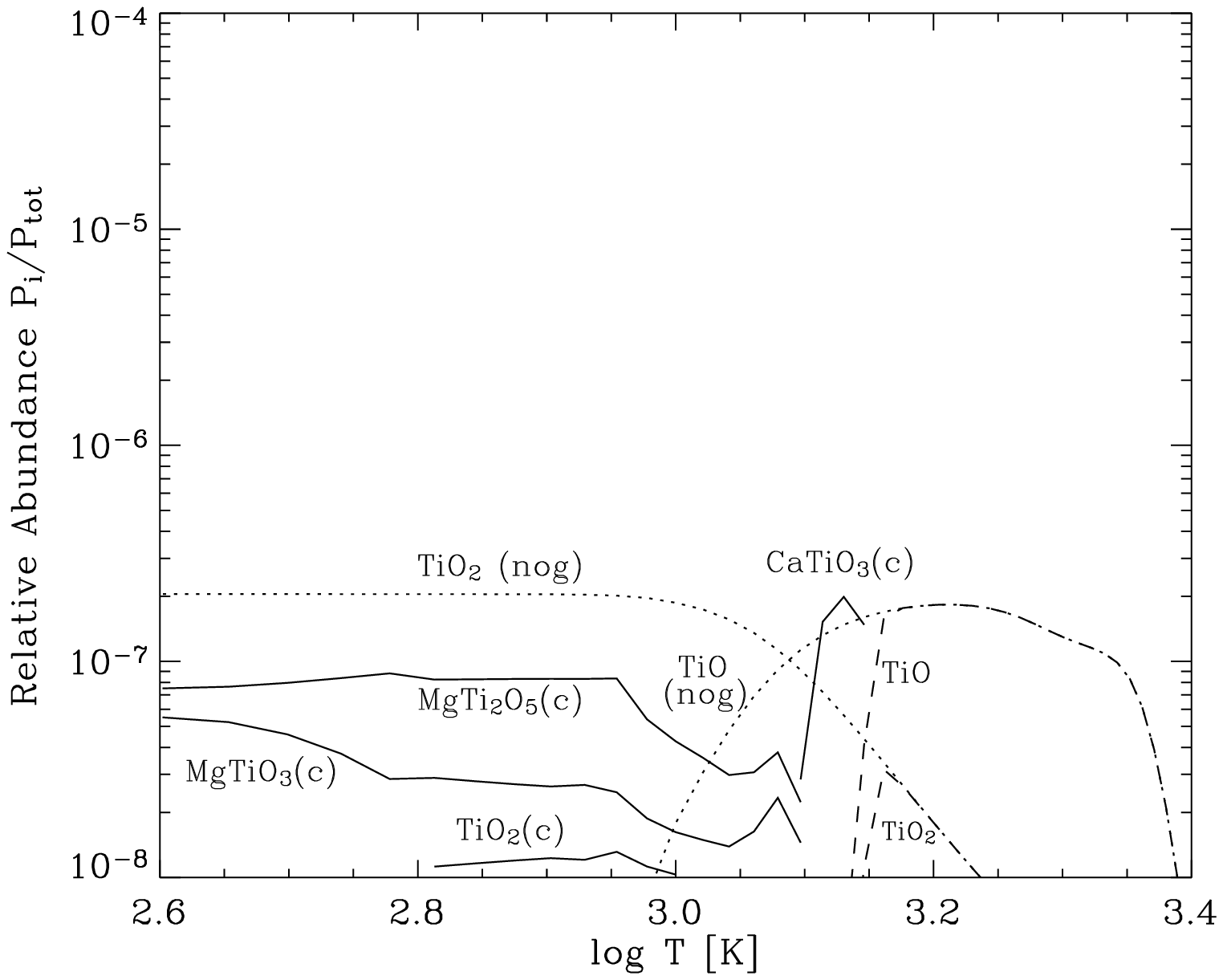}
\caption{The relative abundances of Ti-bearing gas-phase and solid species for a single gas pressure
(1 dyne cm$^{-2}$); the scale is the same as Fig.~1.  The solid lines are the condensates and the 
long dashed lines are molecular TiO and TiO$_2$.  The dotted lines show the abundances of
the molecular species when condensate species are {\em not} included in the calculation.}
\label{Figure 2}
\end{figure}

%%% added () definition of number abundance
Figure~3 shows a contour plot of the relative number abundance (the ratio of the amount of a particular species 
to the total gas density) of several 
species as functions of gas temperature and pressure.  Fig.~1 represents a slice at $\log~P=0$ in Fig.~3.  
For clarity, each contour in Fig.~3
represents a single value of $P_i/P_{gas}$ for each species, and has a different value for each 
species.  The contour values shown are for Al$_2$O$_3$ (all four forms of Al$_2$O$_3$ 
are included) for which the value is -6.2; for CaTiO$_3$ 
the value is -7.0; for MgAl$_2$O$_4$, -6.0; MgSiO$_3$, -5.0; Fe, -4.7; and  
FeS is -5.0.  The species with the highest condensation temperature at $\log~P=0$, Al$_2$O$_3$, forms a ridge
peaking along a line on $\log~T=3.0$ to 3.3 from low to high gas pressure.  Note that at the very highest pressures
the first condensate to form appears to be solid MgAl$_2$O$_4$.  
%%% rewrote sentence for a better explanation
However, because only a single contour value for each species is shown it appears as if Al$_2$O$_3$ does 
not exist at the highest pressures.  This is artificial; Al$_2$O$_3$ still exists, but its abundance 
has fallen below the value of the contour shown.
The species CaTiO$_3$ and MgAl$_2$O$_4$ also form ridges in the $PT-$plane, but very 
little MgAl$_2$O$_4$ forms at low pressures.  
Both MgSiO$_3$ and Fe condense on very broad plateaus: for example, Fe 
forms at temperatures lower than $\log~T=2.9$ at low 
pressures all the way up to $\log~T=3.35$ at the highest pressures.  However, for the condensation
sequence shown here the amount of solid Fe is reduced at $\log~T=2.7$ in favor of solid FeS.

\begin{figure}
\plotone{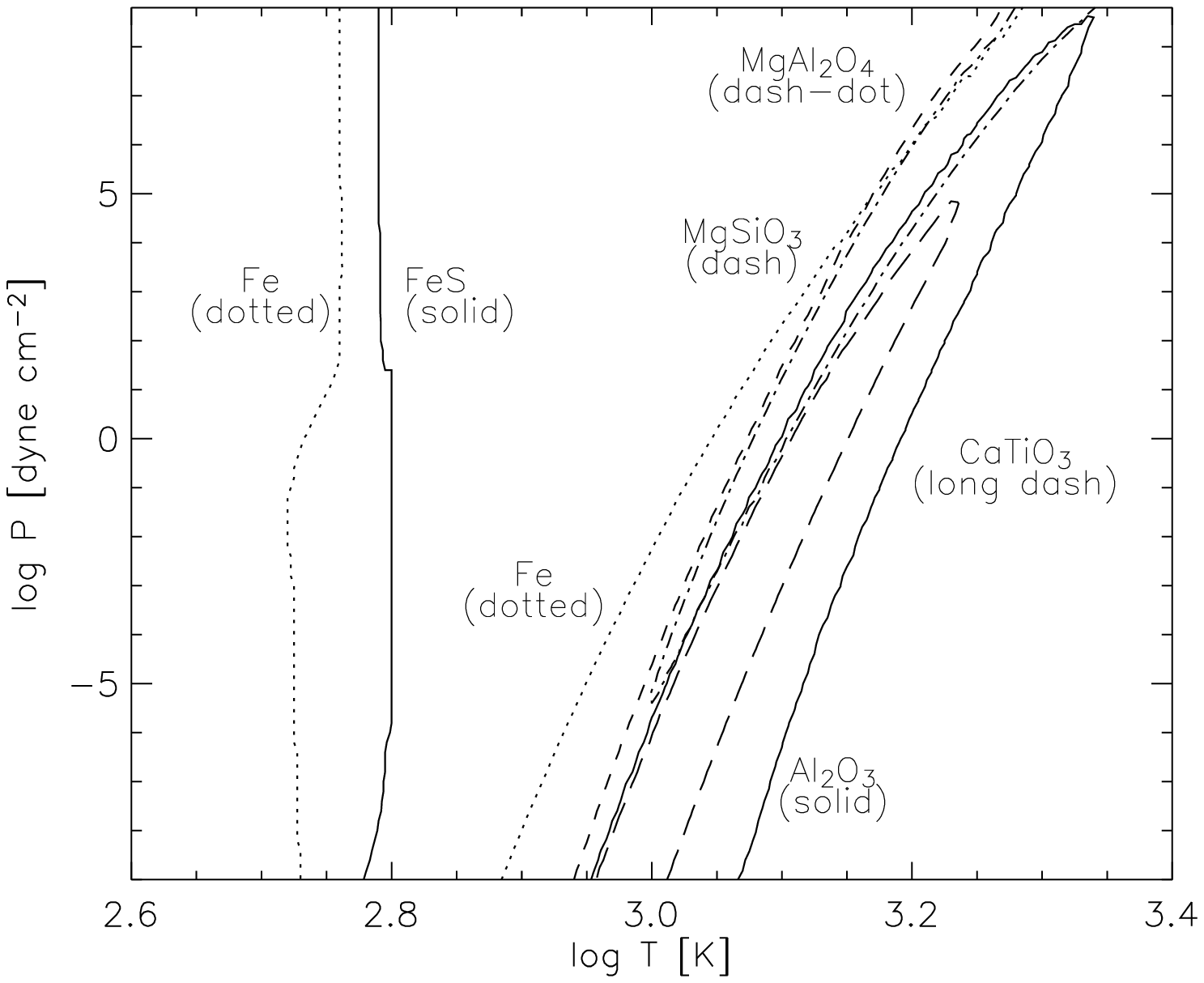}
\caption{A contour plot of the abundance of several grain species as functions
of gas temperature and pressure.  The grains shown are Al$_2$O$_3$ (solid line, at high T), CaTiO$_3$ 
(long dashed line), MgAl$_2$O$_4$ (dash-dot line), MgSiO$_3$ (dashed line), Fe (dotted line), 
and FeS (solid line, at low T).  As described in the text, only one contour value is shown for each species to make
the plot more readable}
\label{Figure 3}
\end{figure}

A contour plot such as the one shown in Fig.~3 is important in understanding how the opacity changes 
%%% added a sentence of explanation
with changes in gas temperature and pressure.  As the gas-grain mixture turns from gas dominated to 
grain dominated the mean opacity changes dramatically: once grains "turn on" the mean opacity will
also depend upon which grain exists and which do not.
The following sections will discussion this relationship between the EOS and the opacity.

\section{Sources of Opacity}
In the following sections we include a discussion of the sources of data for each type 
of opacity included in our calculations.  Discussion of parameters in this section are 
based upon mean opacity tables computed for solar compositions with $3.3 \leq \log~T \leq 
4.3$ and from $-8 \leq \log~R \leq 3$.  The parameter $R$ is defined as 
$\rho / T_6^3$ where $\rho$ is the gas density and temperature is in millions of Kelvin.  

The details of the calculation of the full table of Rosseland mean opacities is described 
in Section~4 below.  Monochromatic opacities are used to compute the Rosseland mean 
($\kappa_R$) by:

\begin {equation}
  \frac{1}{\kappa_R} \equiv \frac{ \int_{0}^{\infty} \frac{1}{\kappa_{\lambda}} \frac{\partial B_{\lambda}}{\partial T} d\lambda }
     {\int_0^\infty \frac{\partial B_{\lambda}}{\partial T} d\lambda }
\end {equation}

%%% rewrote sentence
\noindent where $\kappa_{\lambda}$ is the monochromatic opacity described in this section, $B_{\lambda}$
 is the Planck function and 
$\partial B_{\lambda} / \partial T$ is therefore the weighting function of the Rosseland mean.  
%%% rewrote sentence, deleting 1 of 2
For the computations discussed
here we integrate Eq.~1 over 24,000 wavelengths.  These wavelengths are not evenly spaced, and are 
concentrated between 0.1 and 2.0~$\mu m$: from 10 to 1000 \AA ~we have
points every 10 \AA; from 1000 to 20000~\AA, every 2~\AA; from 20000 to 50000~\AA, every 5~\AA;
from 5~$\mu m$ to 50~$\mu m$, every 100~\AA; and from 50~$\mu m$ to 500~$\mu m$, 
points are spaced every 0.1~$\mu m$.  
%%% added a sentence of explanation
The wavelength points are adjustable and several tests have been made to ensure convergence.  
For example, doubling the number of points increases
the computation time by about a factor of two without significantly changing the 
total Rosseland mean opacity.  Note that this is not the case for Planck mean opacities 
discussed in Section~5.  

\subsection{Continuous Sources}
Descriptions of the continuous opacity sources used in {\sc phoenix} are found 
in \cite{ah95}.  To be complete we list a few of the important sources here.  
For H, He and heavy elemental (C, N, O, Ca, Ti, V, Cr, Mn, Fe, Co and Ni) 
$b-f$ and $f-f$ atomic processes we use the cross-sections of \cite{rm1979}. 
For the other light elements we use the cross-sections of \cite{mathisen1984}.
%%% cleaned up language 
The analytic fits of \cite{verner95} are also included in {\sc phoenix}, 
but they are only significant at very high temperatures, beyond those discussed here.

The negative ions of a few elements can be large sources of $b-f$ and $f-f$ opacity.
For H$^-$ opacity we use \cite{john1988} and for He$^-$ we use a function from 
\cite{vardya1966} which has been modified to reproduce the cross-sections from \cite{john1994}.
For H$_2^-$ we use a formula from \cite{john1975}.  H$_2^+$ among 
other quasi-molecular hydrogen opacity sources are from \cite{cg1969}
and \cite{gingerich1971}. C$^-$ opacities are from \cite{mm1966}.

\subsection{Atomic Line Opacity}
%%% clarified
Table~2 shows the major atomic neutral and ionic species with strong spectal lines 
included in the work presented here.  We 
include 40 atomic species and their ions in our calculations.  
Minor species such as Li, Be, and B among others are included
%%% moved sentence 
in the EOS, but are are not shown in the table 
since their low abundance and low number of lines do not affect mean opacities significantly.  
Since we focus on mean opacities calculated for temperatures less than 30000~K 
we do not include in our calculations ions beyond the 5th ionized stage for any element, but {\sc phoenix} 
does have the capability to compute the opacity due to ions with higher ionized stages. 
Most of the line data is taken from \cite{kuruczCD11993}, with small adjustments for minor species.

Often in model calculations it is convenient to discard weak lines to save computation
time.  The line selection process of {\sc phoenix} is described in \cite{ah95}, but for opacity 
tables computed by our group we include all atomic lines listed in Table~2 
%%% added ()
(plus weaker lines not shown in the table) with lines
%%% broke sentence up
at line center having a line opacity larger than the opacity in the continuum as 
Voigt profiles.  For lines weaker than the continuum we assume Doppler profiles.
\cite{schweitzer1996} contains a discussion of the details of this calculation.

\subsection{Molecular Sources}
Sources for the list of molecular lines used in {\sc phoenix} are described in Table~3 for
many individual molecular sources and in Table~4 for molecular line data from 
the HITRAN database from \cite{hitran1992} and \cite{husson1992}.  As in the computation of the
opacity due to atomic lines, all molecular lines listed are read into {\sc phoenix}; that is
no line selection process is used.  All molecular lines are assumed to have Doppler profiles since
the number of molecular lines per wavelength interval is very large and no molecular lines are strong 
enough to have significant wings. As for the atomic lines \cite{ah95} and \cite{schweitzer1996} 
contain discussions of the details of the calculation.

Water vapor opacity is a very strong source of opacity from about 1800~K to 2500~K.  
As outlined in Table~3 {\sc Phoenix} has the ability to include three different water opacity sources.  
The earlier work of AF94 used the molecular line data from \cite{jj93} and 
%%% added clarification
\cite{jjs1994} for molecular water but this data is not available in our current code. 
For the opacities computed for this paper we have chosen to use the molecular line data 
from \cite{amesh2o1997} commonly used by various authors including \cite{limitdust2001}.  
This list has been shown by \cite{AHS2000} to provide a more complete description of the near-infrared
spectra of brown dwarfs.
%%% removed: despite some remaining discrepancies, especially in hotter stellar atmospheres.

Collision-induced absorption (CIA) opacities are taken from data provided by 
the Borysow group.  A detailed list of the references for the CIA opacities 
can be found in \cite{limitdust2001} and \cite{borysow2001}.

\subsection{Grain Opacity}
Just as is the case for molecules, the computation of the opacity due to small solid 
dust particles requires a knowledge of the abundance of the particles as a function 
of temperature and pressure and the absorption and scattering properties of each particle.  
The opacity due to a dust species can be computed from

\begin {equation}
\kappa_\lambda ~\rho~=~\pi\sum_{i}\int_a n_i(a)~Q_{\rm ext}(a,i,\lambda)a^2~da
\end {equation}

\noindent 
%%% added "normalized", chem. eq. to number abundance, other changes too
%%% this entire paragraph has been rewritten as per referee's comments...
where $n_i(a)$ is the normalized number of dust particles of species {\it i} of size {\it a} 
and $Q_{\rm ext}(a,i,\lambda)$ is the total extinction (absorption plus scattering) 
efficiency of the particle.  The size distribution, $n_i(a)$ depends upon 
both the number abundance
of species {\it i} and the size distribution of dust particles.  
Here the size distribution of dust particles is assumed to be that found for particles in the 
interstellar medium by \cite[hereafter, MRN]{mrn} also assumed by AF94.  The choice of size distribution
is somewhat model dependent and different distributions will change the resulting mean opacities.  
%%%add new sentence
While an interstellar size distribution does not necessarily apply in other physical circumstances,
it has been adopted as a standard by most investigators.  Tables based upon other size distributions
will be computed upon request.
For a good 
review of modern size distributions, see \cite{clayton2003}.  The extinction efficiencies 
of the particles is computed according to Mie theory for solid spheres composed 
of a single, pure substance.  Future work will also explore the effect of different 
size distributions, aggregrate grains, shape, and porosity on the mean opacity of gas-grain mixtures.

Table 5 lists the grain species included in our EOS and for which we have optical 
%%% added reference
data available.  An excellent reference to an online database is found in \cite{jagerjqsrt2003}.
The table gives each condensate by chemical formula, the 
source of the thermodynamic data, source of the optical constants and the wavelength 
range for which the optical data is valid.  The last column in the table indicates if 
the data for the condensate is from an analog species.  
%%% rewrote the following sentences
For example, optical constant data for MgTiO$_3$ is not available, but MgTiO$_3$ exists in the
EOS and has a significant abundance for a narrow range of temperatures (see Fig.~2).  
Optical data for CaTiO$_3$ is available and can be used as an analog species for MgTiO$_3$.  The uncertainty
in whether CaTiO$_3$ is a good analog for MgTiO$_3$ is much smaller than the error in including 
MgTiO$_3$ in the EOS with no opacity from that species.  

%%%redone the entire paragraph...
For the computation of the Rosseland mean opacity, missing optical constant data over specific
wavelength ranges will 
dramatically affect the resulting mean opacity due to the inverse nature of the Rosseland mean (see Fig.~16 of AF94).  
Often optical data is not available over the entire wavelength range that we integrate to compute the Rosseland mean.
To avoid these problems we interpolate 
or extrapolate data so that the optical constant data covers the wavelength range 
from $0.1\mu m\leq \lambda \leq 500\mu m$.  
Care is taken in the extrapolation to follow trends in similar condensate species.  
For example, optical data for Ca$_2$Al$_2$SiO$_7$ is extrapolated into the visual
part of the spectrum by using MgSiO$_3$ and Mg$_2$SiO$_4$ for comparison since 
all three species are silicate in nature and the far-infrared optical constants 
%%% changed errors to uncertainties
are similar in shape.  While such extrapolations undoubtedly introduce uncertainties, however, the 
fact that the Rosseland mean weighting function is so small at short wavelengths (see Fig.~4f) these 
uncertainties are minor.  We stress that Ca$_2$Al$_2$SiO$_7$ 
can be an important condensate at intermediate temperatures (see Fig.~1) and should not be
removed from the mean opacity calculation.

\section{Discussion of Results}

\subsection{Monochromatic Opacity}

Monochromatic opacities of important absorbers are shown in Figure~4
for a range of gas temperature, a constant value of $\log~R = -3$ and solar abundances.  
The important opacity contributors are indicated in each plot. The 
monochromatic opacities in each panel of the figure have been smoothed by convolving
the monochromatic output of {\sc phoenix} with a gaussian.  The width of the 
gaussian varies with wavelength in order to show detail at a wide range of wavelengths.
The wavelength range chosen is based upon the value of the weighting function ($dB/dT$);
each plot is shown where the weighting function has a value greater than 0.1\% of the maximum.
As the temperature of the gas
is decreased the wavelength maxima of the weighting function moves towards longer wavelengths
%%% added equation
as defined by $\lambda_{max}=3600/T$.
It should be noted that the smoothing was only done in preparing these figures 
and is not done in the computation of the mean opacity.

\begin{figure}
\plotone{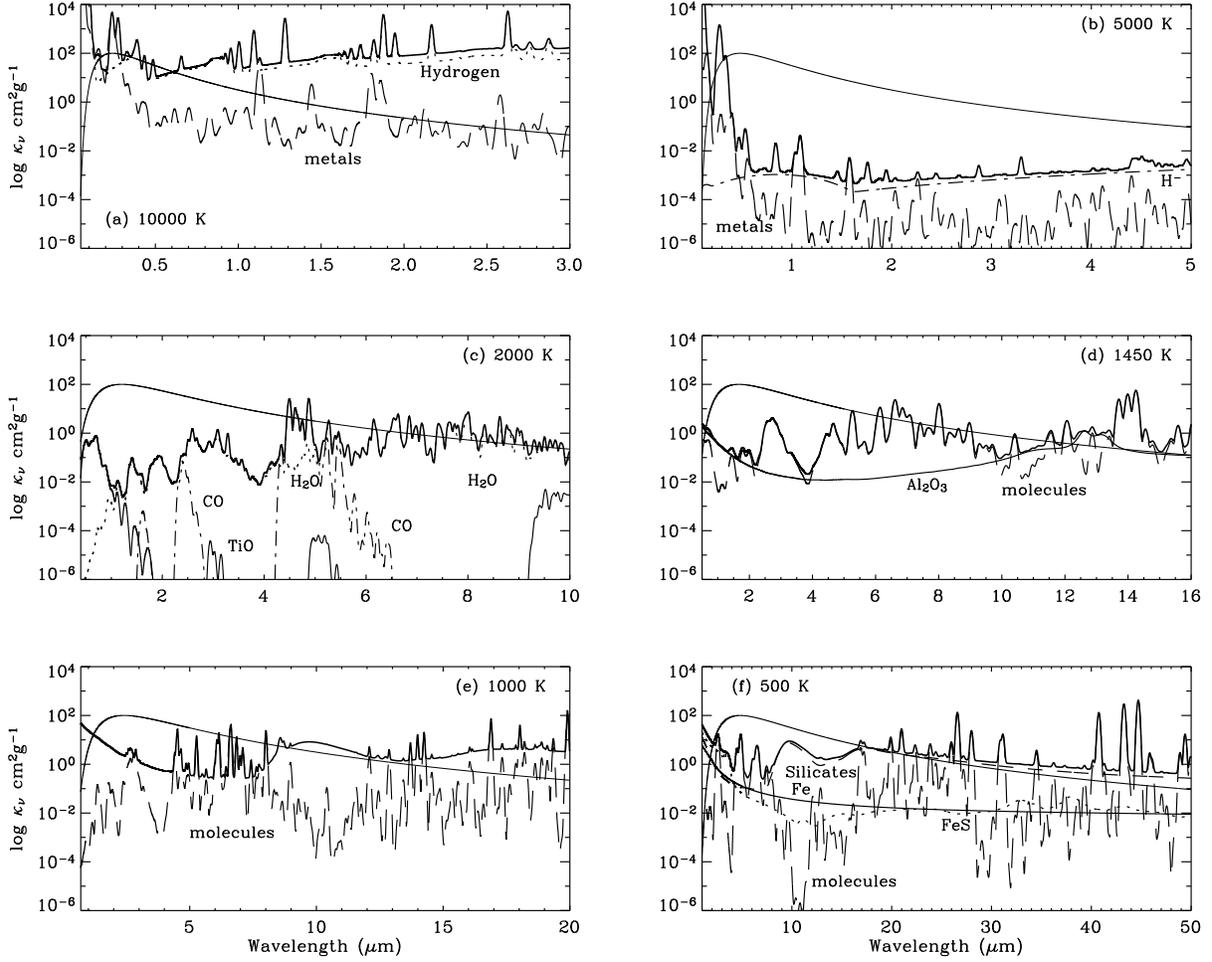}
\caption{The monochromatic opacity at a wide range of temperatures for solar 
abundances and $\log~R = -3$.  Each panel is marked with the representative temperature and 
includes the weighting function, $dB/dT$ (dotted line) 
used in the computation of the mean opacity.  The wavelength range of each plot is
shifted so that the weighting function is within 0.1\% of the maximum and the 
monochromatic opacities were smoothed by convolving the monochromatic output of {\sc phoenix} 
with a gaussian.  See text for a detailed discussion of each panel.}
\label{Figure 4}
\end{figure}

In Figure~4a, at 10000~K, the frequency dependent opacity is dominated by 
hydrogen $b-f$ and $f-f$ absorption with a significant contribution from
atomic line absorbers (labeled ``metals'' in the figure) appearing at short wavelength 
(typically $\lambda < 4000$~\AA).  The atomic line sources become 
even more important at shorter wavelengths, but the value of the weighting function is 
rapidly decreasing.

At 5000~K (panel~b) the contribution of atomic lines increases significantly relative 
to the total opacity, while the opacity due to neutral hydrogen is diminished.   
At wavelengths longer than 0.5~$\mu$m 
H$^-$ is the dominant continuous opacity souce, but is not shown in panel~(a) because it
is not important at 10000~K.  The contribution from molecular sources of 
opacity are not shown in panel~(b) for clarity, but are included in the total opacity.  
%%% added explanation
The bump in the total opacity at 4.5~$\mu$m is due to CO gas, a molecule with high chemical stability.

In panel~(c) of Fig.~4 at 2000~K the monochromatic opacity is beginning to be 
dominated by molecular absorbers.  In the panel, the 
individual contributions from TiO, H$_2$O, and CO are shown.  Between 4000~\AA~and 1.2 $\mu$m,
TiO is the most significant contributer with CO being important at 4.5~$\mu$m.  Water opacity
is the dominant source of opacity over the broad range of wavelengths shown in the panel.  At
2000~K atomic metals are important below 4000~\AA, but are not important to 
the total mean opacity due to the sharp decrease in the weighting function.

At lower temperatures grains begin to become the dominant opacity source. 
Fig.~4d was chosen with a gas temperature of 1450~K (logarithm value 3.16) correlating
to the peak of the Al$_2$O$_3$ abundance shown in Fig.~1.
Fig.~4d shows how important a contribution Al$_2$O$_3$ can make.  
At many wavelengths Al$_2$O$_3$ is the strongest continuous contributor
to the opacity with molecular sources adding a ``forest'' of lines on top of the 
dust source.  Molecular sources still have a large role in the total opacity with 
TiO peaking at wavelengths shortward of 1.2~$\mu$m and H$_2$O contributing at infrared
wavelengths.

At 1000~K the the most abundant grains are the silicates and iron (see Fig.~1).
Fig.~4e shows how important grain sources of opacity can be to the total.
Fig.~4f, with a gas temperature of 500~K, is shown indicating the strength of the individual
contributors: solid Fe, FeS, and Silicate grains 
(for this figure the species MgSiO$_3$, Mg$_2$SiO$_4$, and Fe$_2$SiO$_4$ 
are included as ``Silicates'').

\subsection{Pure Hydrogen Case}
Shown in Figure~5 is the mean opacity from this work compared with OPAL and AF94 for 
the pure hydrogen case for $\log~R = -3, -1, and 1$; pure hydrogen opacities are 
not available from OP.  Results are generally quite good, with differences 
between OPAL and this work being 0.02 dex for the temperatures shown in the plots
for low $\log R$ values.  For the larger $\log~R$ values the difference is 
as large as 0.05 dex or 12\%.  When we remove opacity sources from our 
calculations including all molecular hydrogen species (H$_2$, H$_2^-$, etc.) 
and quasi-molecular hydrogen opacity, then mean opacities at the higher $\log~R$ values 
compare much better with OPAL.
The rise in our opacity at lower temperatures (about 5600~K) is due to our inclusion of 
molecular hydrogen which is not included in OPAL.  

\begin{figure}
\plotone{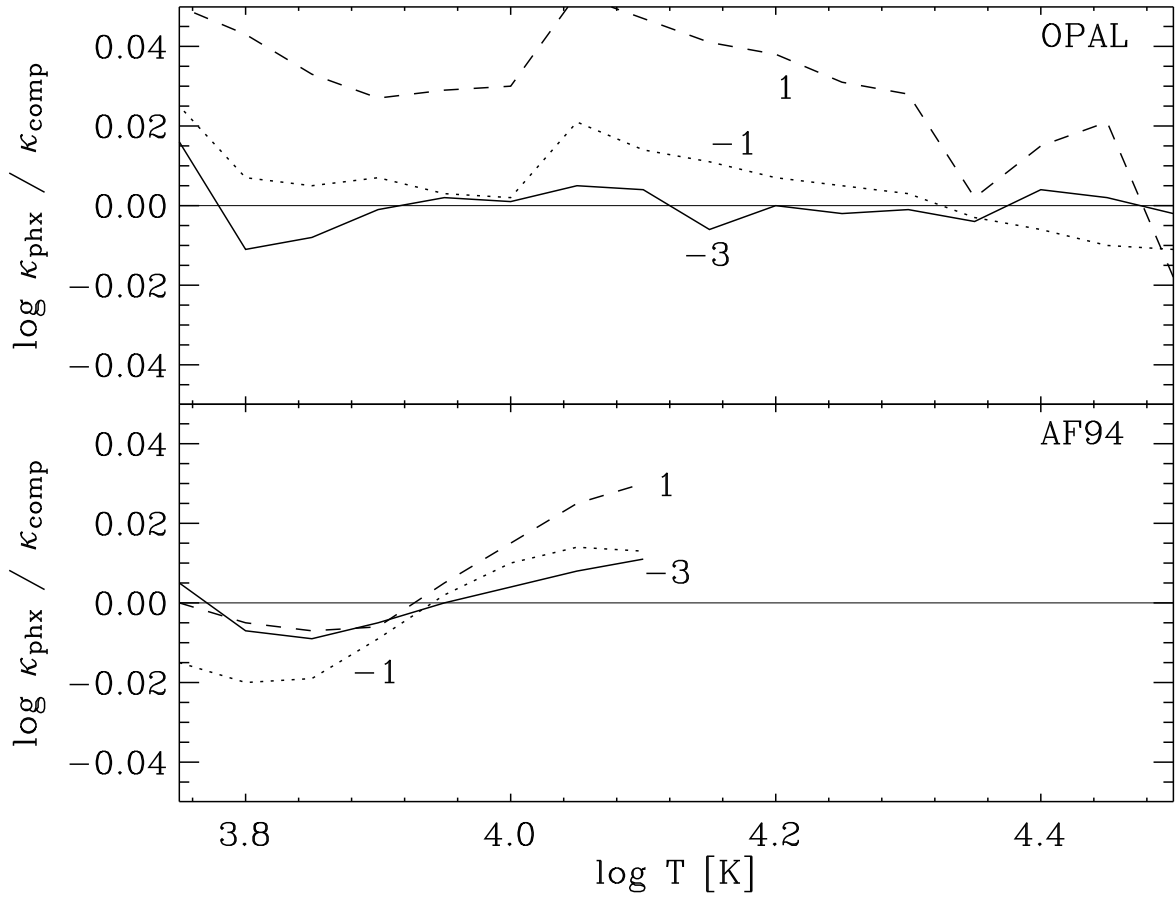}
\caption{Plot of the logarithm of the mean opacity ratio for various values of $\log~R$.  For clarity
each ratio is marked with its value of $\log~R = -3$ indicated as solid lines, -1 as dotted lines
and $\log~R=1$ as dashed lines.  AF94 opacities are not available for $\log~T$ values larger than 4.1.}
\label{Figure 5}
\end{figure}

\subsection{Pure Helium Case}
A comparison between OPAL, OP, and AF94 with present calculations for the pure 
helium case are shown in Figure~6a.  Again the values of $\log~R$ are indicated in each panel.
A large difference is apparent between $3.9 < \log~T < 4.4$ where the mean 
opacity from this paper is much higher than either AF94 and OPAL.  The cause of 
this effect is our inclusion of neutral helium absorption lines which are not 
included by OPAL, OP, or AF94.  Fig.~6b shows this effect with the same ratios as Fig.~6a,
but without neutral He lines included in the computation.  Notice that the scale
of the figure is different than in Fig.~5.  At mid-range temperatures the opacity differences
are similar in absolute magnitude to the pure hydrogen case.

\begin{figure}
\plotone{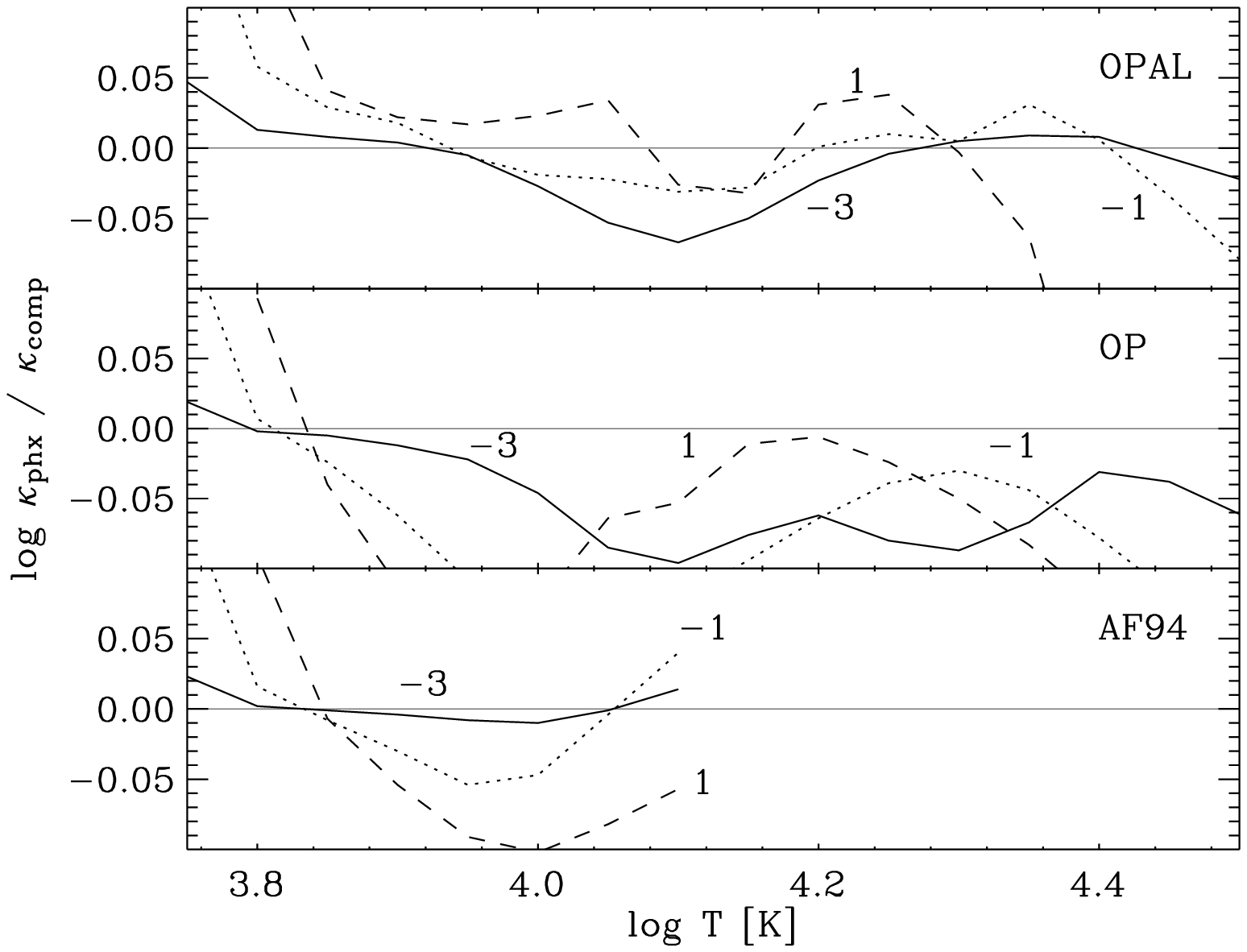}
\caption{The upper panel is the same as Fig.~5 (with the exception of the vertical scale) for the pure He case.  
Large deviations in the opacity ratio are discussed in the text.  The lower panel
is the same as the upper panel except the opacity due to He lines is not included 
in the computations.}
\label{Figure 6a}
\end{figure}
\setcounter{figure}{5}
\begin{figure}
\plotone{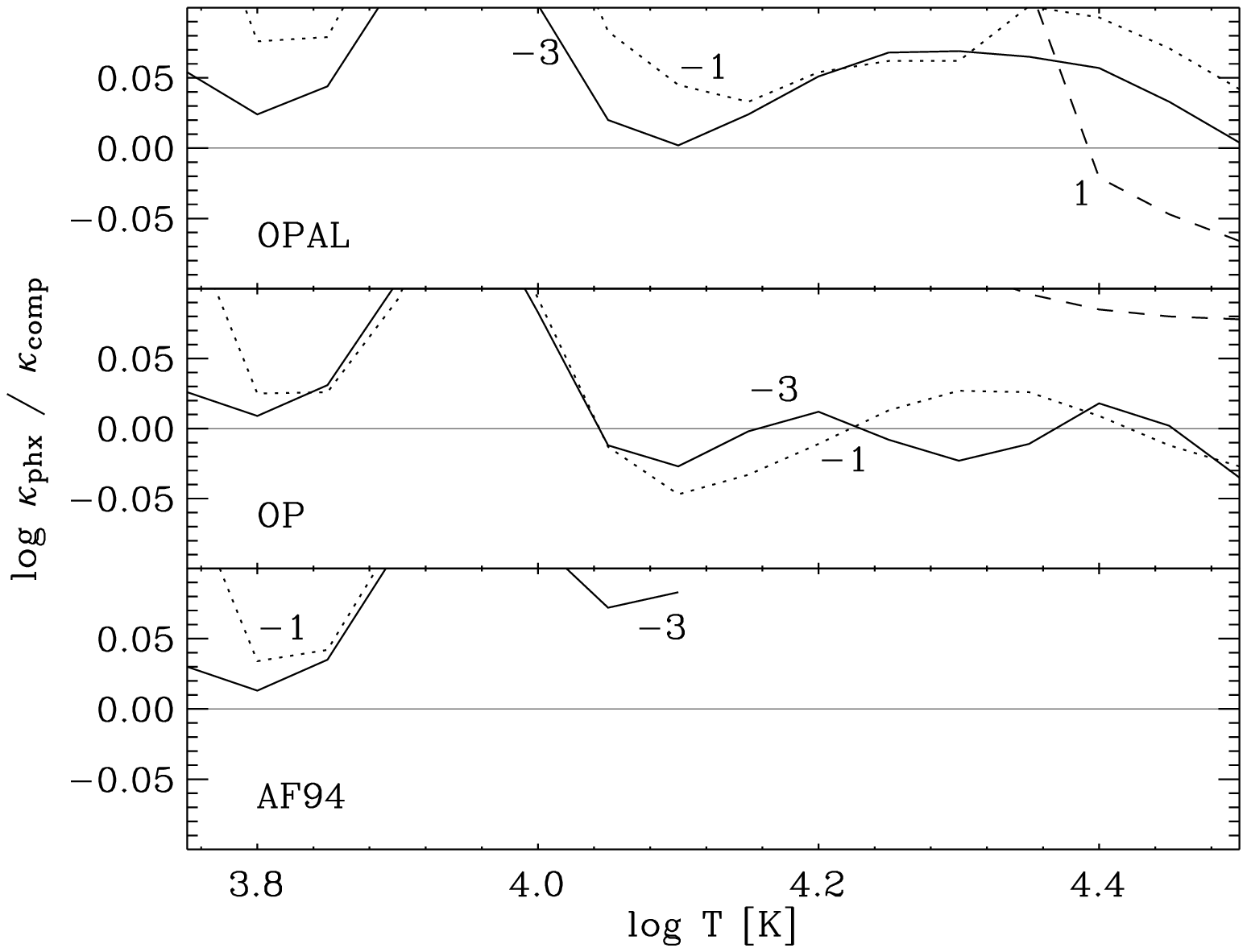}
\caption{Fig. 6b}
\label{Figure 6}
\end{figure}

\subsection{Zero-metallicity Case}
Figure~7 presents the results for a hydrogen and helium mixture, 
specifically $X=0.7$ and $Z=0$.  As in the pure 
hydrogen and the pure helium mixtures the comparisions are good.  
Differences between our opacities and those from OPAL are within 0.03 dex from 
$3.9 < \log~T < 4.3$ for low $\log~R$ values.  
Comparing with OP, differences are as large 
as 0.07 dex in the same temperature range.  This indicates conflicting 
%%% added reference
results between OPAL and OP (see the recent paper by Seaton \& Badnell 2004\nocite{SB2004}).  
Differences increase at lower temperatures 
as molecular hydrogen becomes important in our opacity calculations (see the 
discussion above for a pure hydrogen case).  

\begin{figure}
\plotone{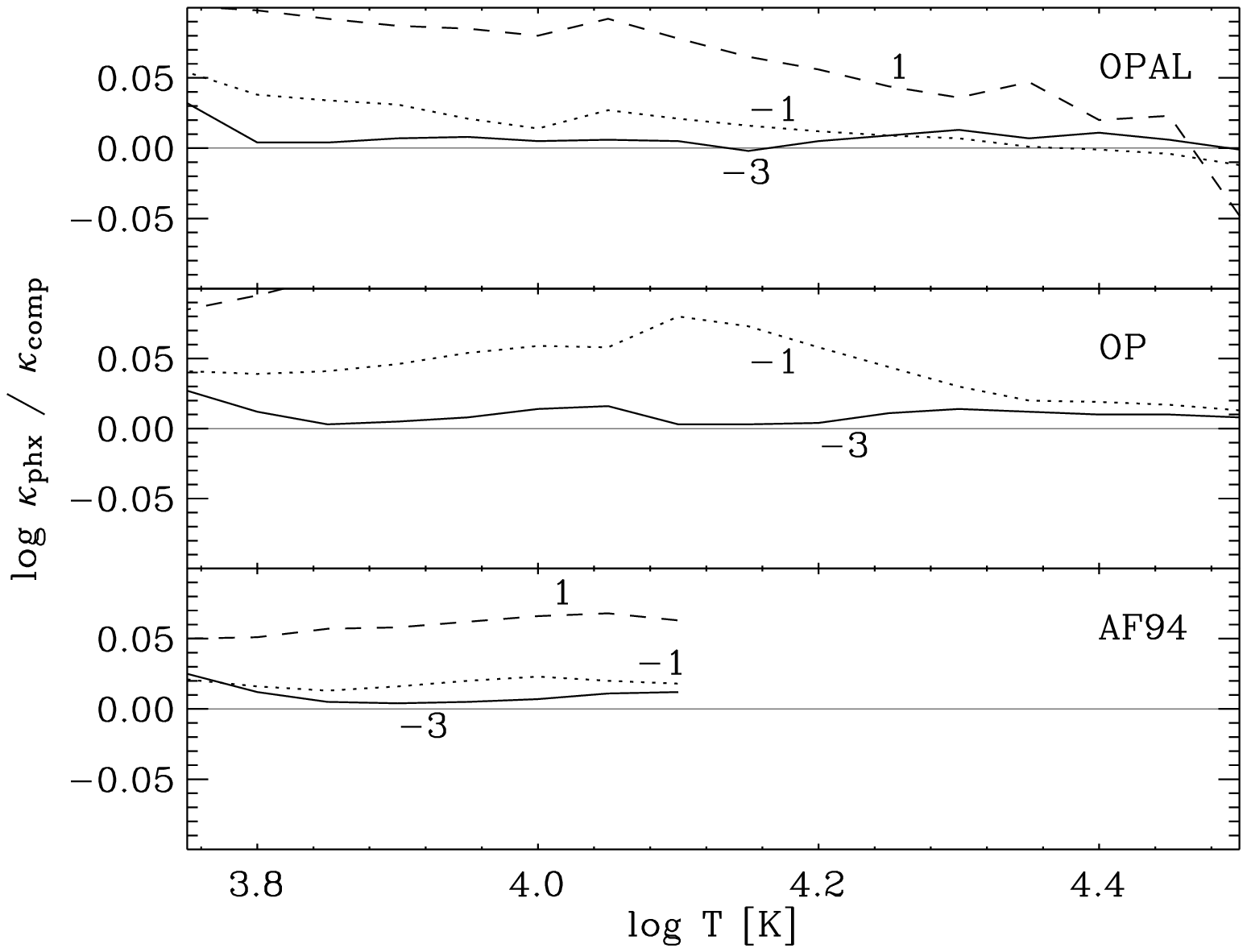}
\caption{Same as Figure~5 for the zero-metallicity case}
\label{Figure 7}
\end{figure}

\subsection{Solar Metallicity Case}
Shown in Figure~8 are comparisons for the solar metallicity case $X=0.7$ and $Z=0.02$ for
the same $\log~R$ values used in previous figures.  
Again, differences are generally less than 0.05 in the logarithm of the mean opacity for 
$\log~T$ values larger than 4.1.  At lower temperatures the effects of molecules on the 
equation of state begin to become important.  

\begin{figure}
\plotone{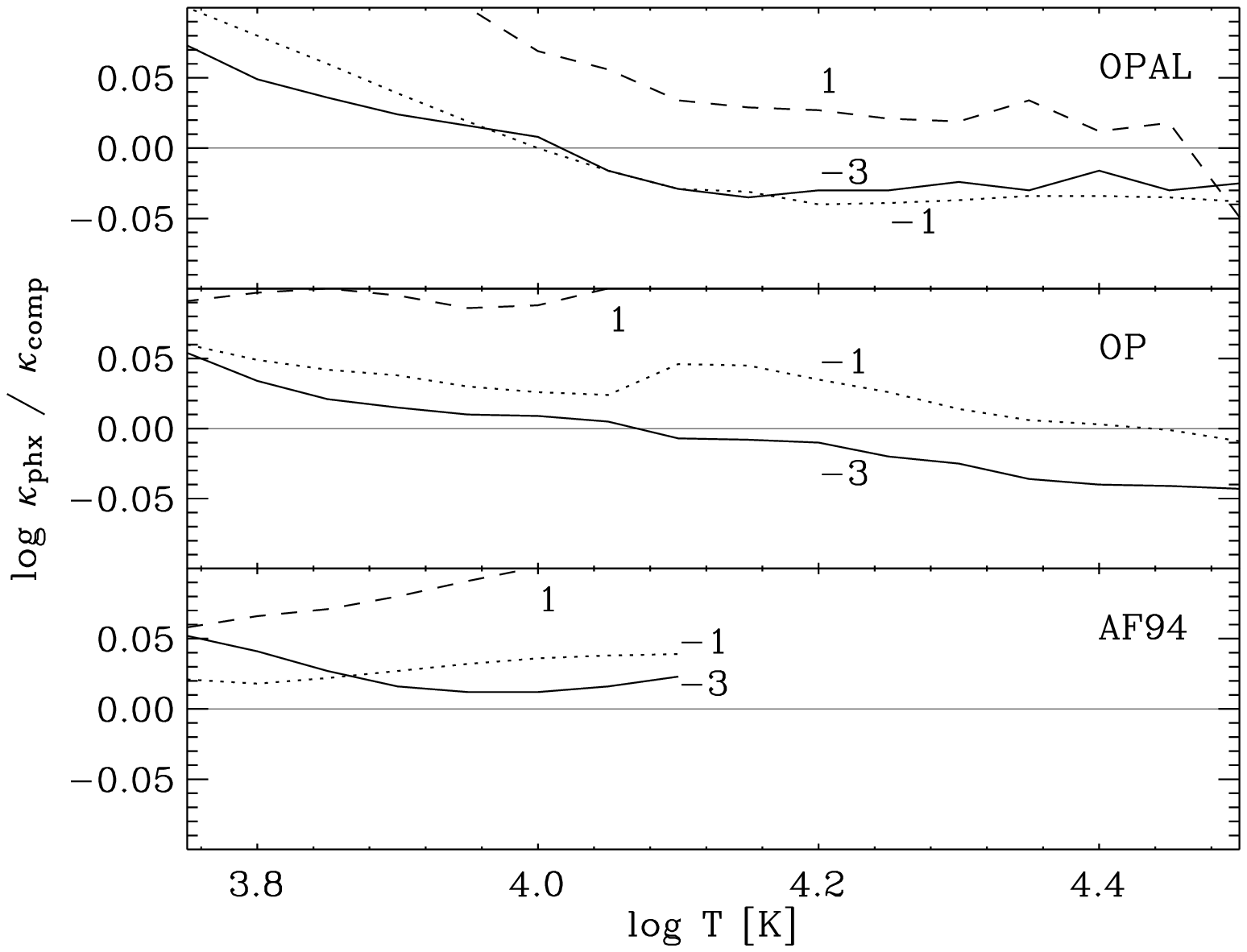}
\caption{Same as Figure~5 for solar metallicity abundances.}
\label{Figure 8}
\end{figure}

%%% added explanation of solar abundances, and changes made to figure
Figure~9 shows the Rosseland mean opacity for solar abundances (X=0.7, Z=0.02 based upon \cite{gn93}), temperatures from 
%%% moved the S03 hereafter to introduction
500~K to 10000~K, and for $\log~R = -3.0$.  Opacities from OP, OPAL and S03
are also shown in the plot.   The features seen in the figure come from the physics discussed 
in Figures~1 (the EOS) and 4 (monochromatic opacity) and the regions where grains, molecules, and atoms
dominate the opacity are also indicated in the figure.
At high temperatures atomic line and continous opacities dominate the Rosseland mean which falls towards
cooler gas temperatures as the atoms become more neutral.  A bump is seen at $\log T \sim 3.6$ where molecules begin to 
become important.  The sharp rise seen to the left of $\log T \sim 3.4$ is due to 
the formation of molecular H$_2$O and TiO (see Fig.~4c). The bump peaks at $\log T \sim 3.3$ and begins to fall
%%% add clarifying phrase
again due to decreased population of excited levels in molecules.  
Grains appear at $\log T \sim 3.16$ (at this gas pressure) as shown in Fig.~1.  For different 
gas pressures the appearance temperature of grains will move cooler at low pressure and move warmer
at higher pressures as shown in Fig.~3.  The appearance or disappearance 
of various grain species account for the rise and fall of the mean opacity towards cooler temperatures in Fig.~9. 
Many of the individual features will be discussed in detail below.

\begin{figure}
\plotone{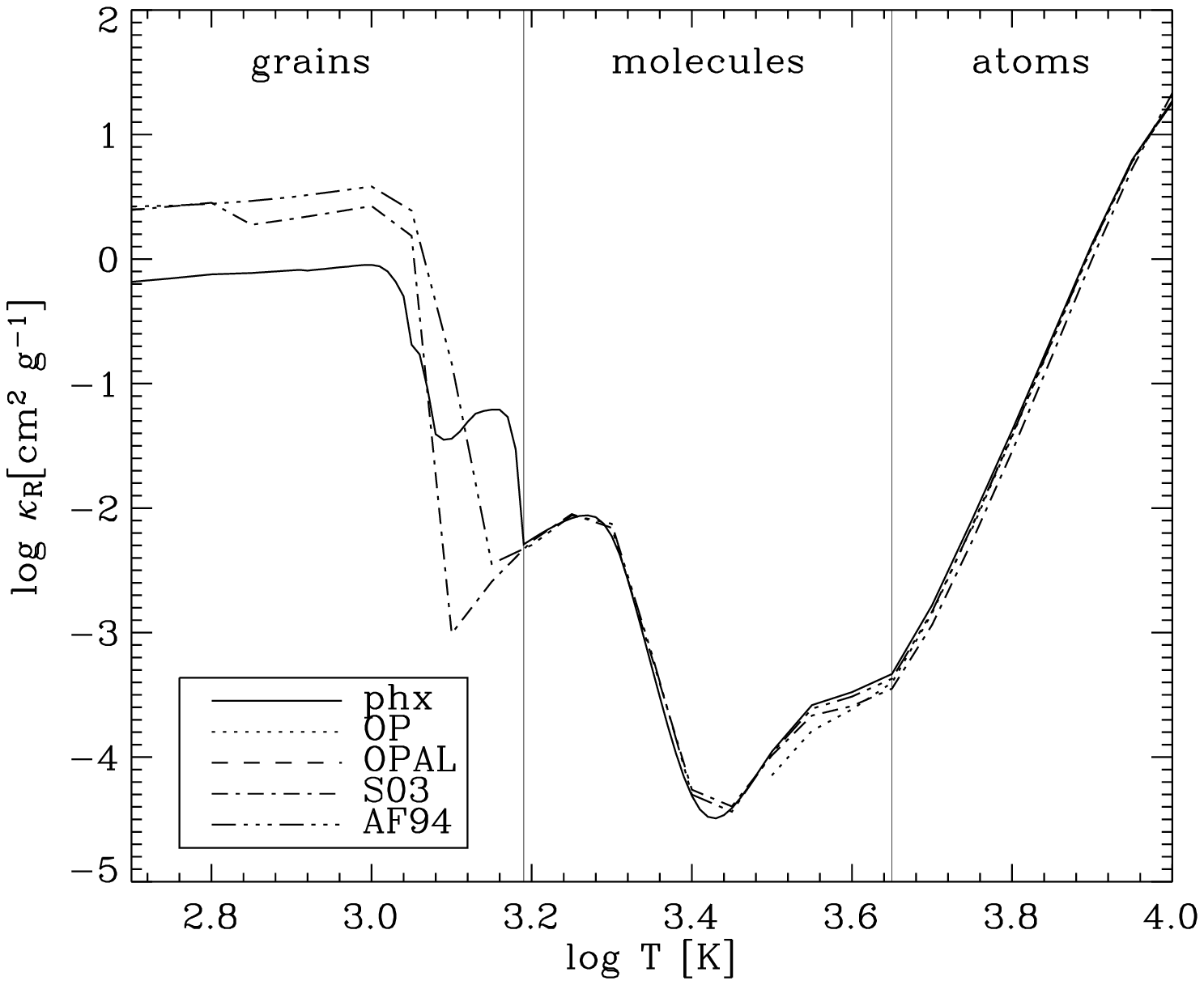}
\caption{Logarithm of the Rosseland mean opacity as a function of temperature for various computation
of the opacity as indicated in the legend.  The compuation is for solar abundances and $\log R~=~-3$.  Regions where
certain species grains, molecules, and atoms dominate the opacity are indicated.
Differences at low temperature are discussed in the text.}
\label{Figure 9}
\end{figure}

Several differences between the sets of opacities are readily apparent, the largest being at low temperature with 
some minor differences in the region dominated by atomic opacities.  
At higher temperatures AF94, OP, OPAL and the current work are not 
resolved in the figure, however, mean opacities from S03 are seen to be somewhat lower than the others.
Fig.~8 shows that
differences between AF94, OP and OPAL are all less than 0.05 in the logarithm.  The difference between the current work and 
S03 is 0.17 dex at $\log T=3.8$, or nearly 50\% lower in value.  Second, the effect of the absence
of molecules in the OP calculation becomes apparent around $\log~T \sim 3.6$.
The dominant opacity source from $\log~T~=~3.25~$to$~3.4$ is line opacity from molecular water and TiO.  Third, the onset
of grain formation, seen at $\log~T~=~3.25$ in Fig.~9, occurs at a lower temperature in AF94, and even lower in S03.
This difference is the result of the more complete gas/grain EOS included in the present work while, as already mentioned,
AF94 only included six grain species in the EOS.  Also both AF94 and S03 do not include high temperature 
condensates such as Al$_2$O$_3$.  In trial computations where
%%% added sentences to explain differences
the higher temperature condensates were not included, the trend shown by S03 was well matched.  Another small factor in the 
differences between this work and S03 is the use of condensation temperature instead of evaporation temperature.  
In a physical circumstance where the temperature is time-dependent, 
these two temperatures do not have the same value due to a hysteresis-like effect:
the evaporation of grain material occurs at a
temperature lower than the condensation temperature, as discussed by S03.  

%%% this para now focuses on AF94 comparisons
A difference of greater concern in Fig.~9 is the magnitude of the Rosseland mean opacity at 
low temperatures when comparing with AF94.
At $\log~T =~3.0$ AF94 has a Rosseland mean opacity that is about half a dex 
greater than present calculations.  There are several reasons for this large difference.  First, the computations of 
AF94 utilized a crude EOS (only six grains species and decoupled grain and gas number densities)
and do not match the number abundances of the current EOS.  Second, AF94 used 
%%% clarified CDE statement
the continuous distribution of ellipsoids (CDE) for computation of the grain opacity which produces
a greater total opacity from a fixed amount of grain material.  Third, AF94 used
optical constants from \cite{Ossenkopf92} for (dirty) astronomical silicates while the current data set is outlined in 
Table~5.  When we change the input 
optical constant data for silicates in {\sc phoenix} to match that of AF94 we come much closer to repeating the 
results of AF94.  Since the EOS of {\sc phoenix} computes the amount of individual species of pure enstatite, forsterite, 
and fayalite we use the optical constants for those pure solids and try not to mix in "astronomical silicates" or 
use data for olivine ((Mg,Fe)$_2$SiO$_4$). 

%%% added new para for disscussion of S03
Comparing present opacities with S03 presents challenges as well.  
The S03 paper includes a variety of models for dust parameters and for Fig.~9 we used the S03 model that
best fit the assumptions of our current calculations, including 
that grains are homogenous spheres with a normal iron content.  Other differences between this work 
and S03 are due to differences in the optical constant data used, 
different EOS methods, and different grain size distributions.  

The importance of the various grain opacity contributors is demonstrated in Figure~10.
In the figure the partial Rosseland mean opacity is shown: that is, the total opacity is computed
excluding one species at a time is plotted.  It is important to point out that in both panels of Fig.~10
the opacity source is removed, but each species still exists in the EOS.  Removing the species
from the EOS can introduce drastic changes to the abundances of other species 
and effect the mean opacity in non-linear ways.

\begin{figure}
\plottwo{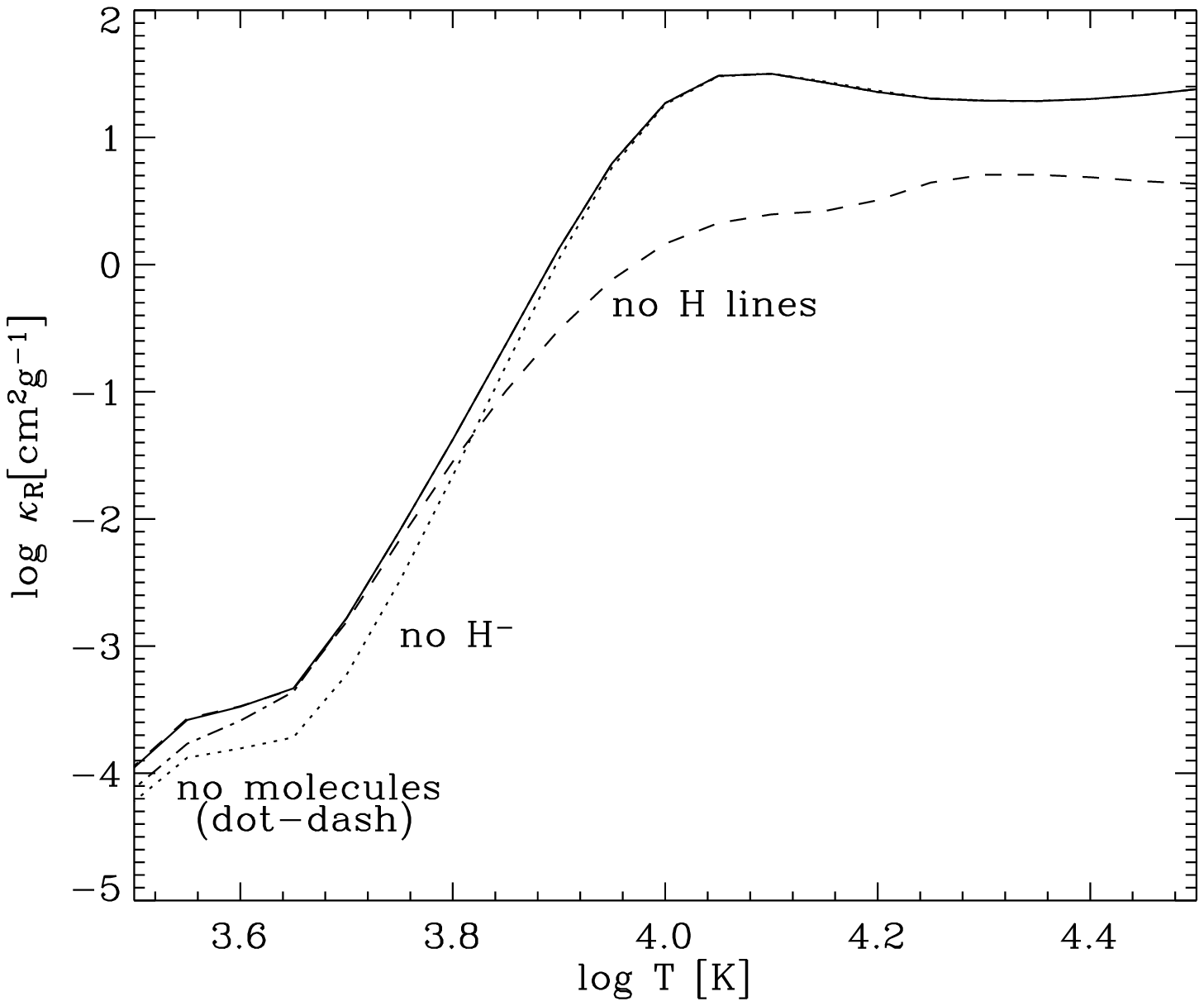}{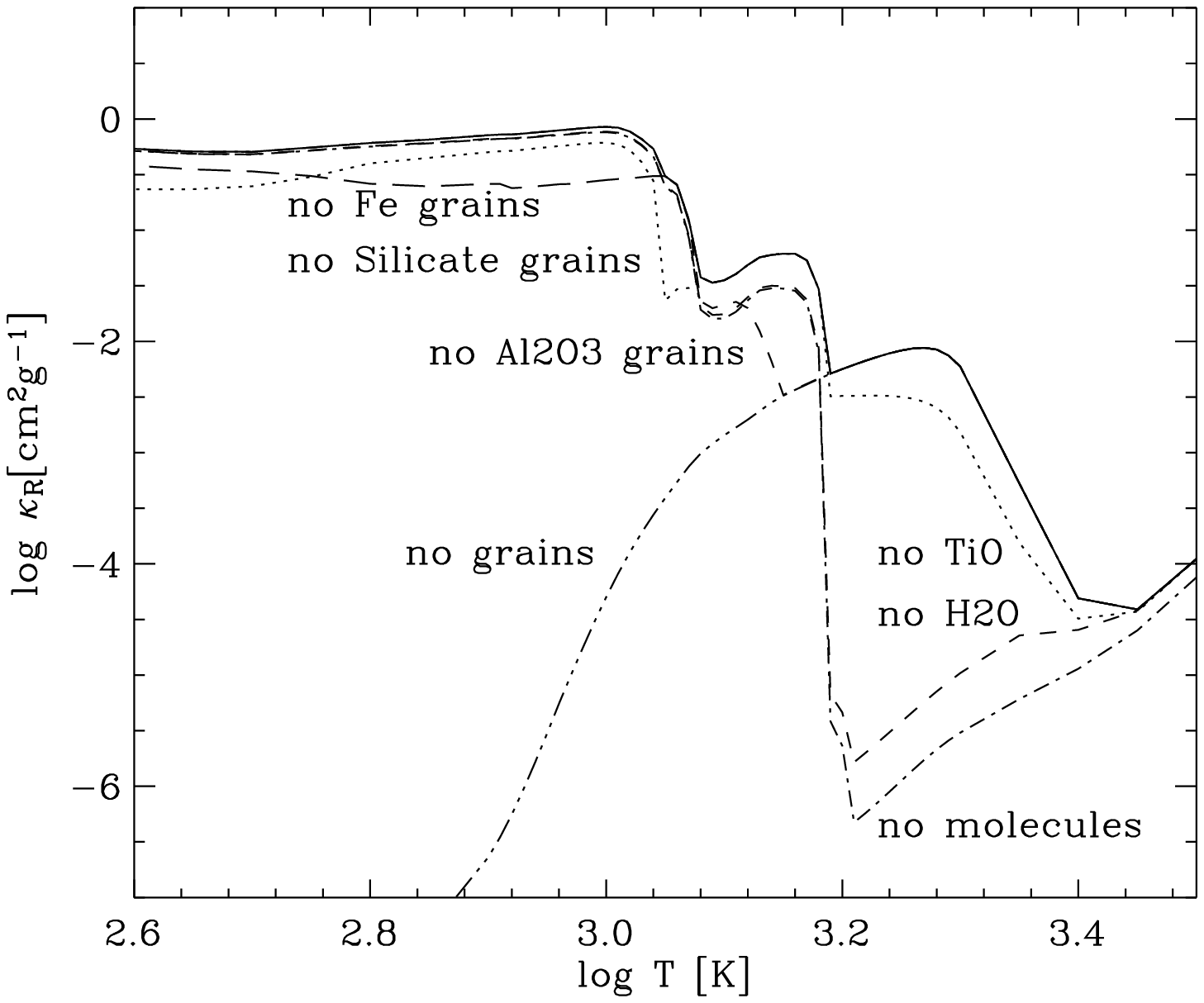}
\caption{Partial Rosseland mean opacity for $\log~R = -3$ with several opacity sources
removed one at a time from the computation.  In the upper figure the high temperature
sources are shown.  The solid line is the total opacity, the dashed is with H~$b-f$ and $f-f$
sources removed, the dotted line removes H$^-$, and the dash-dot line contains no molecules
in the computation.  Similarly in the lower figure each of the lines are marked with the source
that has been removed.  See text for a complete discussion.}
\label{Figure 10}
\end{figure}

In Fig.~10a there are only three dominant opacity sources shown.  Hydrogen contributes the most
at temperatures above 5000~K.  For temperatures less than about 10000~K 
the continuous opacity due to the H$^-$ molecule is 
very important, while below 5000~K molecular sources dominate the mean opacity.  

Lower temperature partial opacities are shown in Fig.~10b.  Again the solid line is the mean opacity
with all species, the dash-dot line removes all molecules from the opacity, the dashed line removes 
only water opacity and the dotted line (only for log T larger than 3.2) removes the contribution of 
molecular TiO.  Notice that removing molecular sources changes the mean opacity by more than
a factor of 1000 at $\log~T~=~3.3$.  At even lower temperatures removing all grain species 
(see the dash-dot-dot-dot line in Fig.~10b) from the opacity has a dramatic effect on the 
mean opacity, for example at $\log~T~=~2.8$ the difference is more than six orders of magnitude.  
In Fig.~10b the dotted line (below $\log~T~=~3.1$) is the opacity without silicate grain sources,
and the long dashed line is the opacity without condensed iron.

\subsection{Metallicity Dependence}
Figure~11 shows the effect on the opacity due to changes in metallicity.  For $\log~R=-3$ 
changes in the amount of metals as a function of Z are shown in Fig.~11a and changes with 
amount of hydrogen (X) are shown in Fig.~11b. For the variable Z case, notice that not only
is the total opacity diminished as the amount of metals is reduced, but that condensation
%%% added explanatory statement
temperatures are also reduced since there are less metals available for grains to exist in
equilibrium.  For Z=.1 the onset of grains occurs at about $\log~T=3.2$,
but for a highly reduced metallicity, Z=.0001, the onset occurs at about $\log~T=3.1$, more 
than 300~K cooler than the higher Z value.  Even though the amount of metals has been reduced by 
three orders of the magnitude at these gas temperatures the amount of the Rosseland opacity has
been reduced by more than four orders of magnitude.  When grains exist they are a powerful opacity source.

%%% added explanatory statement
In Fig.~11b with constant Z and variable X the grain formation temperatures are
about the same since hydrogen is in such great abundance and does not play a role in the grain chemistry.  
The amount of opacity seen at the higher temperatures is greatly affected
however, as the hydrogen fraction is reduced and the abundance of helium is raised especially
at X=.1 and helium lines dominate the opacity at gas temperatures larger than 15000~K.  The molecular
bump seen at $\log T \sim 3.6$ (Fig.~9) is slightly reduced as is the rise due to molecular H$_2$O
at $\log T \sim 3.4$ since there is less hydrogen available.  
The rises in the opacity due to condensates is slightly affected with the formation
%%% added "to form water"
temperatures ($\log T \sim 3.2$) being slightly higher with less hydrogen to form water.

\begin{figure}
\plottwo{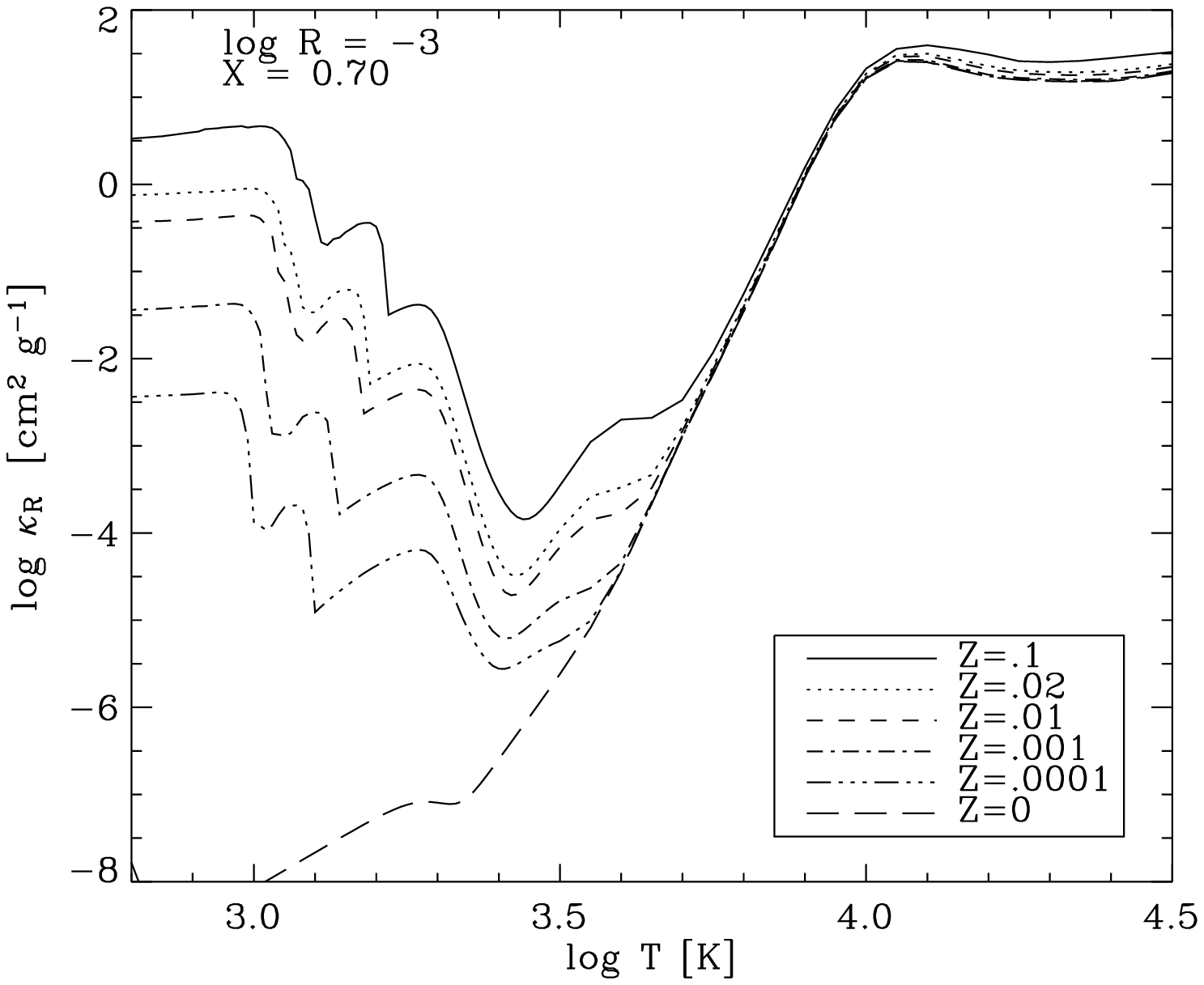}{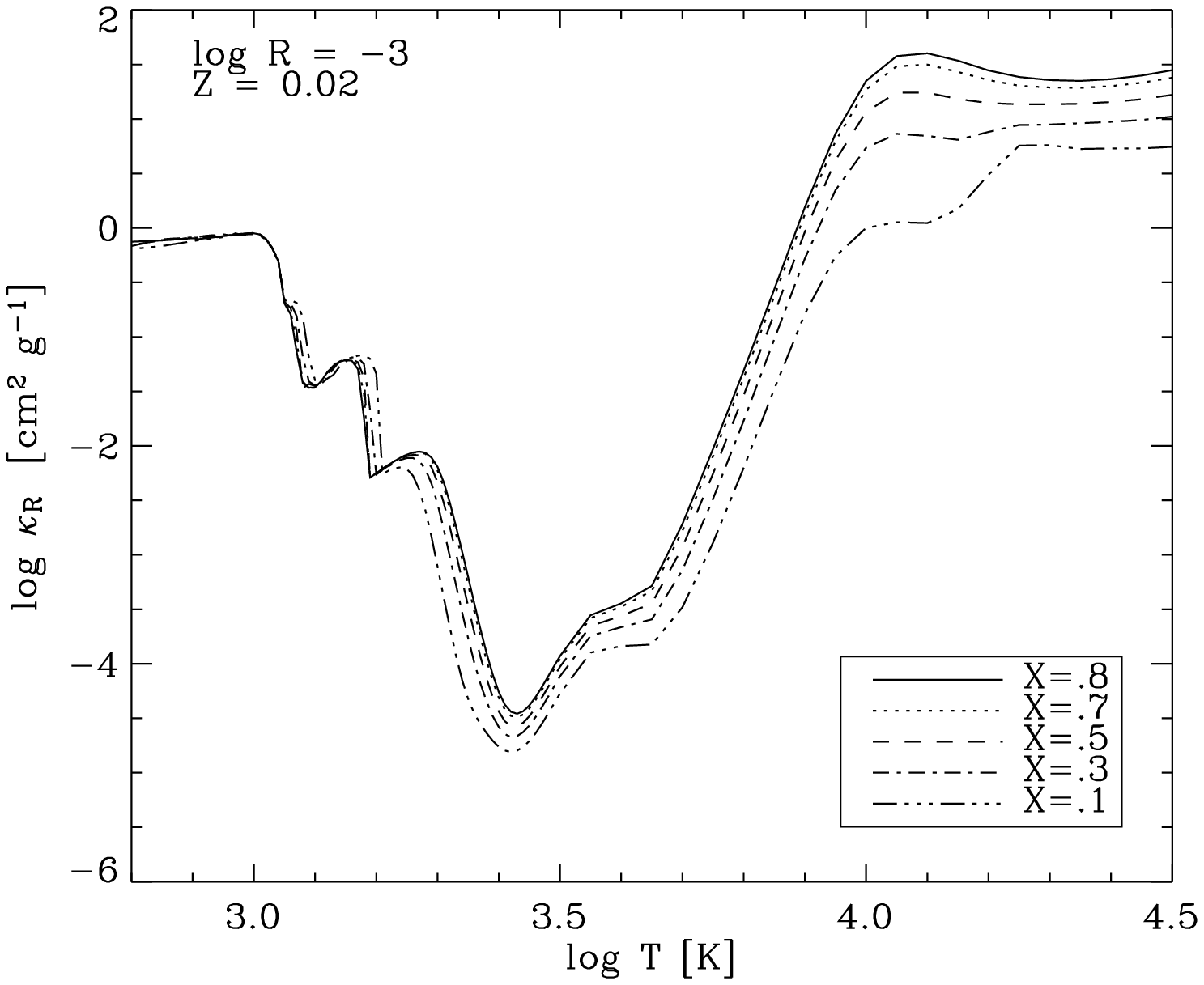}
\caption{Rosseland mean opacity as a function of metallicity (panel a) and hydrogen fraction (panel b).
In panel (a) lines refer to the total opacity with X=0.7, held fixed, and with the values of Z indicated in the legend.  
In panel (b) the lines refer to the total opacity with Z=0.02, held fixed, and with X indicated in the legend.}
\label{Figure 11}
\end{figure}

\section{Planck Mean Opacities}
The Rosseland mean opacity is often used in the limit where the diffusion approximation is a
valid description of the physical conditions of a gas, typically in optically thick regions.  In 
optically thin regions the diffusion approximation may not be a valid approximation. 
In such cases the Planck mean opacity ($\kappa_{P}$) can be used to 
represent the mean opacity and is defined as

\begin {equation}
  \kappa_P \equiv \frac{ \int_0^\infty {\kappa_{\lambda}} B_{\lambda} d\lambda }
     {\int_0^\infty B_{\lambda} d\lambda }
\end {equation}

\noindent where variables are defined as in Eq.~1.  The main difference between a Rosseland and Planck
mean opacity is the way in which the monochromatic opacity is averaged.  In the case of a Rosseland mean,
the inverse of the opacity, multiplied by $\partial B/\partial T$, is summed making the total mean a harmonic mean, 
that is, the mean 
heavily weights wavelengths of low opacity, such as gaps between lines.  
For the Planck mean the monochromatic opacities, multiplied by the Planck function, are
summed directly, so that the total mean Planck opacity is a representation of the opaqueness of
the gas rather than its transparency.

Figure~12 shows a comparison between our calculations and a representation of calculations from the literature
(including OP, S03, and \cite{sharp92}).
Note that the figure is for a single gas density of 10$^{-10}$~g~cm$^{-3}$ and not in $\log~R$ space.
Also included in Fig.~12 is a {\sc Phoenix} computation with more than one million wavelength points,
more than 40 times the normal wavelength 
set as defined in Section~3.  The normal wavelength set of 24,600 wavelengths is not sufficient to 
match the computation from OP.  It is the opacity sampling method of atomic lines (and molecular lines at lower
temperatures) employed in our calculation that causes this defficiency.  This effect is fully expected due to the way
in which OP computes the Planck mean opacity; they essentially sum the oscillator strengths for each line 
(Seaton, private communication) removing the uncertainty of line profile shape and wing cutoffs.

What is puzzling about Fig.~12 is the nearly five order of magnitude difference between the 
Planck opacities of OP and S03 at high temperatures where atomic line sources dominate the mean opacity.  
At low temperatures when the opacity of the gas is dominated by continuous dust sources the results are very similar.  
AF94 opacities are high due to the use of the CDE for the calculation of the grain opacity.  
We can account for differences at high temperature between our work, AF94 and OP because of 
our use of opacity sampling for line opacities and the number of wavelengths points used.  
%%% replaced sentence
It is speculated that S03 and \cite{sharp92} are different from the current work due to differences in the 
molecular line lists used, EOS and number of wavelengths over which the mean opacity is integrated.

The final point we wish to make is that users of AF94 Planck opacities need to strongly consider using our 
newer Planck opacities computed with over one million wavelength points. The newer calculations nicely overlap
with OP at higher temperatures and we are confident of their validity as shown in Fig.~12.

\begin{figure}
\plotone{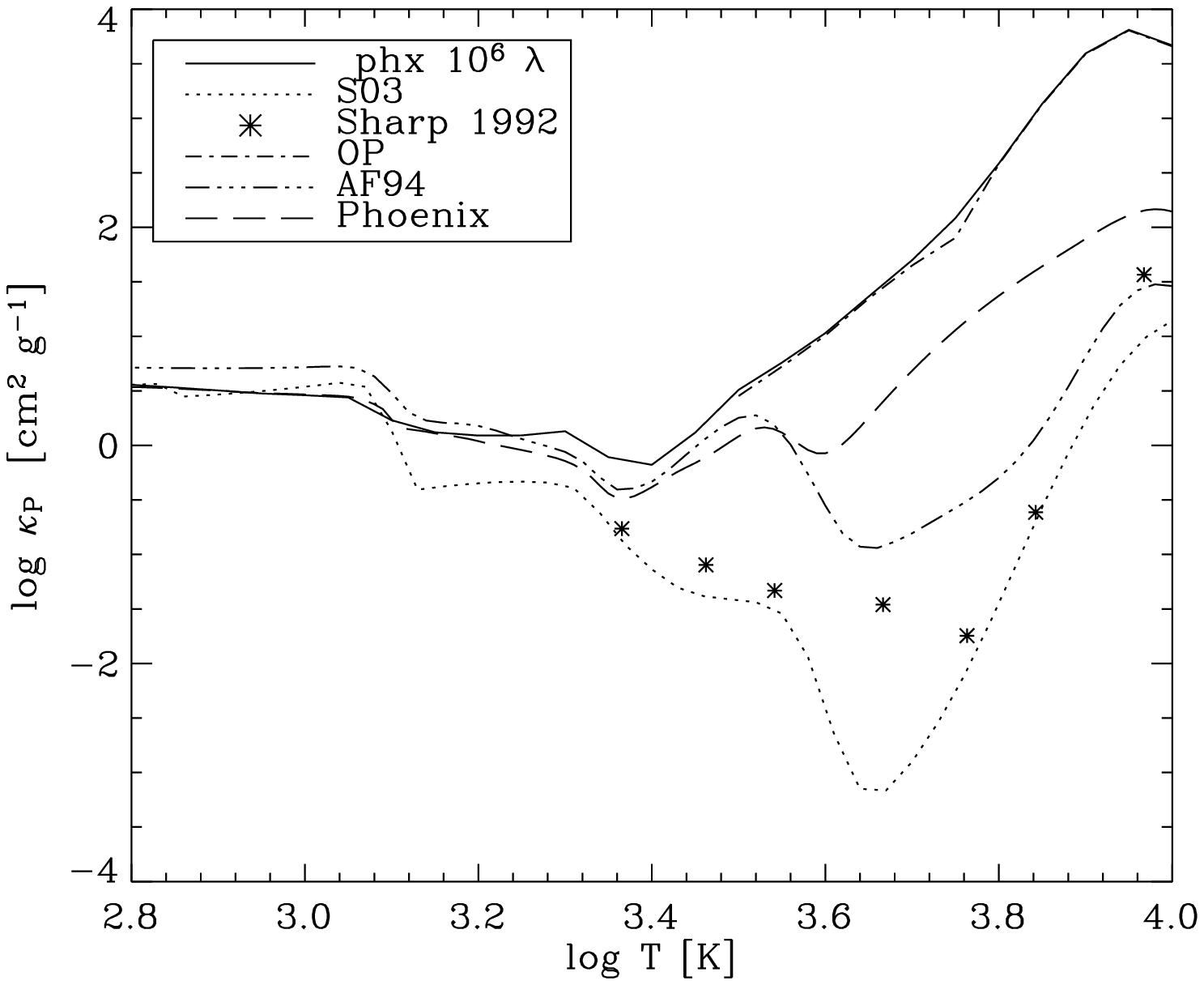}
\caption{Total mean Planck opacity as a function of temperature for a single gas density of 10$^{-10}$~g~cm$^{-3}$
The solid line is a present calculation with over 1 million wavelength points in the wavelength integral.  
The long-dashed line is from the normal 24599 wavelength calculation
discussed in the text.  The dash-dot line is data from OP and the dash-dot-dot-dot line is from AF94.
The dotted line is data from \cite{semenov2003} and the stars are from \cite{sharp92}.}
\label{Figure 12}
\end{figure}

\section{New Opacity Tables}

Many updates have been made to the previous WSU low temperature opacity tables.  
AF94 was only valid below 12000~K, while current tables are valid up to 30000~K and there is better agreement 
with OP and OPAL at high temperatures than AF94.  The present work includes many more wavelength points in
the calculation of the mean opacity and includes a greater number of molecular lines for most sources than AF94, most 
of which are based upon improved laboratory observations and quantum mechanical calculations. 
For dust species, the EOS of the dust is now fully coupled to the gas.  Many more dust species are also included in
the current work.

The standard set of opacity tables that we compute range from $2.7 \leq \log~T \leq 
4.5$ in 0.05 dex increments above $\log~T>3.5$ and below $\log~T<2.9$.  Between those
temperatures the stepsize is 0.01 dex in order to gain resolution of the 
discontinuities at the grain formation boundaries.  For the density parameter {\em R} 
we compute tables with a range $-8 \leq \log~R \leq 1$ in steps of 0.5 dex.  Each table contains
75 temperature and 19 density points for a total of 1425 computations.

For chemical abundances, our base mix is based upon \cite{gn93} and is then scaled
to X= 0.0, 0.1, 0.2, 0.35, 0.5, 0.7, 0.8, 0.9 or Z= 0.0, 0.00001, 0.00003, 0.0001, 0.0003, 0.001, 0.002,
0.004, 0.01, 0.02, 0.03, 0.04, 0.06, 0.08, 0.1 for a total of 120 abundance tables in each set.  

Opacity tables for the base set are available for download at our web site available at 
{\em http://webs.wichita.edu/physics/opacity}.  Custom abundance sets are
%%%added density possibility
available upon request as are tables in density space rather than $\log~R$ as well.

\acknowledgments
Low temperature astrophysics at Wichita State University is supported by 
NSF grant AST-0239590, NASA LTSA grant NAG5-3435,
NASA EPSCoR grant NCC5-168 and NSF EPS-9874732 
with matching support from the State of Kansas.  We are most grateful to the
referee, D. Semenov, for insightful comments on the manuscript.  
PHH was supported in part by the P\^ole Scientifique de Mod\'elisation 
Num\'erique at ENS-Lyon. Some of the calculations presented here were 
performed at the H\"ochstleistungs Rechenzentrum Nord (HLRN), at the National 
Energy Research Supercomputer Center (NERSC), supported by the U.S. DOE, and 
at the San Diego Supercomputer Center (SDSC), supported by the NSF. We thank 
all these institutions for a generous allocation of computer time.  
FA acknowledges support from the PNPS and PNP CNRS programs, as well as to computing resources provided 
by CINES and IDRIS in France.  
We also acknowledge the support from the National Science Foundation under 
Grant No. EIA-0216178 and Grant No. EPS-0236913, matching support from the 
State of Kansas and the Wichita State University High Performance Computing Center. 

\newpage
%\bibliography{ms}

%
% FIGURE CAPTIONS
% Finally, we have figure captions.  Usually these must be on a separate
% page, although unlike table, it is often permissible to have several
% figure captions on the same page.  We force the page break between
% the reference list and the figure captions.

% TABLES are next
\clearpage

\begin{deluxetable}{lllll}
%\rotate 
\tablecolumns{5} 
\tablewidth{0pc} 
\tablecaption{Improvements in Low Temperature Opacity Calculations} 
\tablehead{ 
\colhead{} & \colhead{A75\tablenotemark{a}} & \colhead{AJR\tablenotemark{b}}   & \colhead{AF94\tablenotemark{c}}    & \colhead{Present work} }

\startdata
Equation & super-     & decoupled gas  & decoupled gas  & gas \& dust\\
of State & saturation & \& dust        & \& dust        &  in equilibrium\\
         & ratio      &                &                &  \\
\\
Molecular & straight& $2 \times 10^5$ lines +   & $3 \times 10^7$ lines & $8 \times 10^8$ lines \\
 Opacity & mean  & straight mean \\
& & water\\
\\
Dust Opacity & 	1 species & 3 species & 6 species & 31 species\\
& Rayleigh & Mie & CDE & Mie \\
\\
Number of & 50 & 900 & 9000 & 24000+\\
Wavelengths\\
\enddata

\tablenotetext{a}{\cite{a75}} \tablenotetext{b}{\cite{ajr}} \tablenotetext{c}{\cite{af94b}}

\end{deluxetable} 
\clearpage

\begin{deluxetable}{llllll}
%\rotate 
\tablecolumns{6} 
\tablewidth{0pc} 
\tablecaption{Number of atomic lines for each ion} 
\tablehead{ 
\colhead{Element} & \colhead{I} & \colhead{II}  & \colhead{III} & \colhead{IV} & \colhead{V} }

\startdata
H       & 457   &      &     & & \\
He      & 1630  & 463  &     & & \\
C       & 8885  & 979  & 493 & 201  & \\
N       & 13858 & 1725 & 402 & 499  & 233\\
O       & 2311  & 3954 & 1114& 543  & 529\\
F       & 2363  & 2164 & 4919& 889  & 273\\
Ne      & 4177  & 9119 & 860 & 838  & 220\\
Na      & 385   & 205  & 708 & 195  & 313\\
Mg      & 2668  & 580  & 846 & 1211 & 248\\
Al      & 869   & 3184 & 369 & 171  & 132\\
Si      & 8526  & 919  & 1539& 326  & 131\\
P       & 2413  & 1213 & 163 & 214  & 221\\
S       & 834   & 1118 & 338 & 73   & 42\\
Cl      & 6982  & 1676 & 974 & 228  & 71\\
Ar      & 3838  & 10452& 1726 & 160  & 80\\
K       & 702   & 67   & 633 & 111  & 153\\
Ca      & 12860 & 946  & 11740 & 82222 & 330004\\
Sc      & 191270 & 49811 & 1578 & 16985 & 130563\\
Ti      & 897313 & 264874 & 23742 & 5079  & 37610\\
V       & 1156793 & 925330 & 284003 & 61630 & 8427\\
Cr      & 434743 & 1304043 & 990951 & 366851  & 73222\\
Mn      & 327762 & 878996 & 1589314 & 1033926 & 450293\\
Fe      & 789192 & 1264969 & 1604934 & 1776984 & 1008385\\
Co      & 546132 & 1048188 & 2198940 & 1569347 & 2032402\\
Ni      & 149926 & 404556 & 1309729 & 1918070  & 1971819\\
\enddata

\end{deluxetable} 

\clearpage

\begin{deluxetable}{llll}
%\rotate 
\tablecolumns{4} 
\tablewidth{0pc} 
\tablecaption{Thermodynamic and spectral line data for molecules} 
\tablehead{ 
\colhead{Molecule} & \colhead{Source of}     & \colhead{Number of}  & \colhead{Source of} \\ 
 \colhead{}        & \colhead{Thermo. data}  & \colhead{lines}      & \colhead{line data} }  

\startdata
H$_2$O    & 1   & 349,074,613 & 7  \\
          &     & 101,455,142 & 8  \\
          &     & 6,139,497   & 1  \\
TiO       & 2  & 174,027,629  & 9  \\
CH$_4$    & 3  & 11,854,112  & 10  \\

H$_3$$^+$ & 4  & 3,070,572   & 11  \\
CN        & 2  & 2,245,378   & 12  \\
SiO       & 2  & 1,429,165   & 13  \\
ZrO       & 2  & 265,724     & 14  \\
CO        & 2  & 134,421     & 15,16,17  \\

MgH       & 2  & 162,621     & 18,19  \\
VO        & 2  & 7,182       & 20  \\
CrH       & 5  & 2,670       & 20  \\
FeH       & 2  & 2,158       & 21  \\
YO        & 2  & 975         & 22  \\

C$_2$     & 2  & 3,458,871   & 19  \\
SiH       & 2  & 77,642      & 19  \\
CH        & 2  & 71,569      & 19  \\
NH        & 2  & 36,163      & 19  \\
H$_2$     & 6  & 28,486      & 19  \\

OH        & 2  & 26,349      & 19  \\
\enddata

{\em References.-}
(1) \cite{mt94};
(2) \cite{hh79};
(3) \cite{janaf};
(4) \cite{h3p1995};
(5) \cite{tsuji1973};
(6) \cite{saumon93};
(7) \cite{amesh2o1997};
(8) \cite{scanh2o2001};
(9) \cite{AHS2000};
(10) \cite{hahbSTDS};
(11) \cite{H3+1996};
(12) \cite{jl1990};
(13) D. Carbon 1995, private communication;
(14) \cite{littleton1985};
(15) \cite{goorvitch1994};
(16) \cite{gc1994a};
(17) \cite{gc1994b};
(18) \cite{weck2003};
(19) \cite{kuruczCD11993};
(20) R. Freedman 1999, private communication;
(21) \cite{pd1993};
(22) J. Littleton 1987, private communication.

\end{deluxetable} 

\clearpage
\begin{deluxetable}{llll}
%\rotate 
\tablecolumns{4} 
\tablewidth{0pc} 
\tablecaption{Thermodynamic and spectral line data for minor molecules} 
\tablehead{ 
\colhead{Molecule} & \colhead{Source of}     & \colhead{Number of lines}  & \colhead{Source of} \\ 
 \colhead{}        & \colhead{Thermo. data}  & \colhead{}                 & \colhead{line data} }  

\startdata

O$_3$   & 1   & 164,359     & 4  \\
HNO$_3$ & 1   & 117,476     & 4  \\
CO$_2$  & 2   & 59,883      & 4  \\
NO$_2$  & 1   & 55,406      & 4  \\
SO$_2$  & 2   & 26,225      & 4  \\
N$_2$O  & 1   & 23,812      & 4  \\
HOCl    & 1   & 13,300      & 4  \\
OH      & 3   & 8,671       & 4  \\
NO      & 3   & 7,319       & 4  \\
CH$_3$Cl & 2  & 6,687       & 4  \\
ClO     & 1   & 5,966       & 4  \\
NH$_3$  & 2   & 5,787       & 4  \\
H$_2$O$_2$ & 1 & 5,444       & 4  \\
H$_2$CO & 1    & 2,701       & 4  \\
PH$_3$  & 2    & 2,886       & 4  \\
C$_2$H$_2$ & 2 & 1,258     & 4  \\
OCS     & 2    & 737         & 4  \\
HCN     & 2    & 772          & 4  \\
HCl     & 3    & 371         & 4  \\
N$_2$   & 3    & 120         & 4  \\
HF      & 3    & 107         & 4  \\
\enddata

{\em References.-}
(1) \cite{janaf};
(2) \cite{irwin88};
(3) \cite{hh79} and \cite{rosen70};
(4) \cite{hitran1992} and \cite{husson1992}.

\end{deluxetable}

\clearpage

\begin{deluxetable}{llllll}
%\rotate 
\tablecolumns{6} 
\tablewidth{0pc} 
\tablecaption{Thermodynamic and optical constant data for condensates} 
\tablehead{ 
\colhead{Condensate} & \colhead{Common} & \colhead{Source of}  & \colhead{Source of} &  \colhead{$\lambda$ range} & \colhead{Analog}\\ 
\colhead{}           & \colhead{name}   & \colhead{thermo.}    & \colhead{optical}   &                            &  \\
\colhead{}           & \colhead{}       & \colhead{data}       & \colhead{constants} &                            &     }  

\startdata
$\alpha$-Al$_2$O$_3$  & Corundum & 1    & 4     & $0.21\mu \leq \lambda \leq 55.6\mu$  &   \\
$\gamma$-Al$_2$O$_3$  & Sapphire & 1    & 5      & $0.2\mu \leq \lambda \leq 400\mu$    &   \\
                      &          &      & 6   & $1\mu \leq \lambda \leq 7\mu$        &   \\
C                     & Carbon   & 2    & 7  & $0.139\mu \leq \lambda \leq 300.42\mu$   &   \\
CaMgSi$_2$O$_6$       & Diopside & 2    &                  &                                      & from Ca$_2$Al$_2$SiO$_7$ \\

Ca$_2$Al$_2$SiO$_7$   & Gehlenite& 2    & 8   & $6.69\mu \leq \lambda \leq 852\mu$   &   \\
Ca$_2$MgSi$_2$O$_7$   & Akermanite& 2   &                 &                                      & from Ca$_2$Al$_2$SiO$_7$ \\
CaSiO$_3$             & Wollasnonite & 2&              &                                      & from MgSiO$_3$ \\
Ca$_2$SiO$_4$         & Larnite  & 2    &                  &                                      &   from Mg$_2$SiO$_4$ \\
CaTiO$_3$             & Perovskite& 2   & 9  & $0.1\mu \leq \lambda \leq 1000\mu$   &   \\                 

Cu                    & Copper   & 1    & 10      & $0.517\mu \leq \lambda \leq 55.6\mu$    &   \\
Fe                    & Iron     & 1    & 11      & $0.2\mu \leq \lambda \leq 285.7\mu$     &   \\
FeS                   & Troilite & 1    & 12      & $0.1\mu \leq \lambda \leq 500\mu$       &   \\
Fe$_2$SiO$_4$         & Fayalite & 2    & 13      & $2.\mu \leq \lambda \leq 10000\mu$      &   \\
Fe$_2$O$_3$           & Hematite & 1    & 14      & $0.21\mu \leq \lambda \leq 55.6\mu$     &   \\

Fe$_3$O$_4$           & Magnetite& 1    & 14      & $0.21\mu \leq \lambda \leq 55.6\mu$     &   \\
H$_2$O (ice)          & Water ice& 3    & 15      & $2.5\mu \leq \lambda \leq 200\mu$       &   \\
                      &          &      & 16      & $0.0992\mu \leq \lambda \leq 10000\mu$  &   \\
H$_2$O (water)        & Water liquid & 3& 14      & $1\mu \leq \lambda \leq 10^7\mu$        &   \\
MgAl$_2$O$_4$         & Spinel   & 1    & 17      & $1\mu \leq \lambda \leq 300\mu$         &   \\

MgSiO$_3$             & Enstatite & 1    & 18     & $0.1\mu \leq \lambda \leq 624\mu$   & \\
Mg$_2$SiO$_4$         & Forsterite& 1    & 18     & $0.1\mu \leq \lambda \leq 948\mu$   &   \\
MgTiO$_3$             & Geikeilite& 1    &        &                                     & from CaTiO$_3$  \\
MnS                   & Alabandite& 1    &        &                                     &  from FeS \\
NH$_3$                & Ammonia ice & 3  & 19     & $1\mu \leq \lambda \leq 300\mu$     &   \\

NaCl                  & Salt     & 1     & 20     & $1\mu \leq \lambda \leq 300\mu$     &   \\
Ni                    & Nickel   & 1     & 21     & $0.667\mu \leq \lambda \leq 285.71\mu$     &   \\
Nb                    & Niobium  & 1     & 22     & $1.24$\AA$ \leq \lambda \leq 10.33\mu$     &   \\
SiC                   & Moissanite & 1   & 23     & $0.1\mu \leq \lambda \leq 243.673\mu$     &   \\
$\alpha-$SiO$_2$      & Silicon dioxide  & 1 & 24 & $6.67\mu \leq \lambda \leq 487.4\mu$     &   \\

Ti                    & Titanium & 1      & 11     & $0.667\mu \leq \lambda \leq 200\mu$     &   \\
TiO$_2$               & Titanium dioxide  & 1  & 25 & $1\mu \leq \lambda \leq 300\mu$     &   \\
ZrO$_2$               & Zirconium dioxide & 1  & 26 & $4.545\mu \leq \lambda \leq 95.238\mu$     &   \\

\enddata

{\em References.-}
(1) \cite{janaf};
(2) \cite{sh90};
(3) \cite{carlson1987};
(4) \cite{querry1985};
(5) \cite{koike1995};
(6) \cite{Begemann1997};
(7) \cite{rm1991};
(8) \cite{Mutschke1998};
(9) \cite{posch2003};
(10) \cite{ordal1985};
(11) \cite{ordal1988};
(12) \cite{egan1977};
(13) \cite{fabian2001};
(14) \cite{querry1985};
(15) \cite{hudgins1993};
(16) \cite{pollack1994};
(17) \cite{tt1991};
(18) \cite{jager2003};
(19) \cite{nh3opac1984};
(20) \cite{ep1985};
(21) \cite{ordal1987};
(22) \cite{lh1991};
(23) \cite{pegourie1988};
(24) \cite{philipp1985};
(25) \cite{rib1985};
(26) \cite{dowling1977}.

\end{deluxetable}

\end{document}